# Evaluating AI models' capability to automate voice phishing attacks

Fred Heiding [a,*], Claudio Mayrink Verdun [b], Simon Lermen [c], Andrew Kao [a], Vitor Albiero [d], Lauren Deason [d], Irina-Elena Veliche [d], Christine Lehane [d]

[a] *Harvard Kennedy School, 79 John F. Kennedy St, Cambridge, MA, 02138, US*
[b] *Harvard School of Engineering and Applied Sciences, 150 Western Ave, Allston, MA, 02134, US*
[c] *Independent Researcher*
[d] *Meta Platforms, Inc., 1 Hacker Wy, Menlo Park, CA, 94025, US*

ABSTRACT

Voice phishing (vishing) attacks have traditionally been limited by the need for human operators. The rapid emergence of high-quality AI voice synthesis and large language models (LLMs) reduces this bottleneck and enables scalable, automated scams. In this paper, we conduct a large-scale survey experiment (N = 4100) and qualitative interviews (N = 12) to assess U.S. adults' susceptibility to AI-powered voice phishing attacks. Participants were exposed to audio recordings or transcripts of scam scenarios generated using leading voice models such as Llama Full Duplex (Llama FD), Sesame, Gemini, OAI AVM, Play.AI, and ElevenLabs and the corresponding human baselines. The results show high compliance rates. Up to 36% of participants would or might comply with phishing requests in the "relative-in-distress" category. Overall compliance rate across all five scam categories was 16.5%, a striking figure given the low cost and high scalability of AI-automated voice phishing. Caller persuasiveness was the strongest predictor of compliance and certain models (most notably Sesame) achieved ratings comparable to human voices, or sometimes even slightly surpassing them. Our economic analysis suggests that while human-operated vishing is unprofitable at US wages, AI-powered vishing appears to be economically viable for several models. The primary risk of present-day AI-enabled vishing thus lies in the economics of automation rather than novel or "superhuman" persuasive techniques, though these cannot be ruled out for future systems. This raises significant concerns for the design of AI systems, consumer protection, and model release policies.

## 1. Introduction

The human voice carries a unique persuasive power that is difficult to replicate with text alone. When listening to speech, individuals automatically process vocal cues such as prosody, intensity, timing, and emotional coloration, which play a central role in rapid social inference and judgments of authenticity and trustworthiness (Belin et al., 2017; McAleer et al., 2014). In contrast, text-based channels omit many of the nonverbal signals that support credibility assessment in spoken interaction (Kiesler et al., 1984). Our qualitative interviews further support this asymmetry, as several participants reported that vocal cues in AI-generated calls made them more suspicious of voice-based scams, whereas the same conversational content presented as text appeared less overtly artificial and therefore less suspicious. This asymmetry makes voice a particularly powerful medium for social engineering. Scammers have long exploited this vulnerability through voice phishing, or vishing, a form of social engineering that leverages real-time conversations to extract sensitive information or make the recipient take other harmful actions. Unlike email phishing (Tabassum et al., 2024), which can be automated and distributed to millions at negligible cost, vishing has historically required a human operator for each call, limiting its scalability. These bottlenecks ensured that the threat remained bounded by the labor costs of recruiting and training live callers for different languages.

AI systems are now removing these constraints. Recent advances in AI, particularly in LLMs, have increased the sophistication and scalability of social engineering attacks (Heiding & Lermen, 2025; Heiding et al., 2024). The convergence of LLMs and realistic voice generation has created the technical foundation for fully automated, real-time, contextually appropriate voice phishing attacks at unprecedented scale (Figueiredo et al., 2024; Pias et al., 2024). Voice cloning technologies create further problems, as a few seconds of audio is often enough to replicate a specific person's voice

* Corresponding author.
*E-mail addresses:* fred@heiding.io (F. Heiding), claudioverdun@g.harvard.edu (C. Mayrink Verdun), simon.lermen@alumni.tu-berlin.de (S. Lermen), andrewkao@g.harvard.edu (A. Kao), valbiero@meta.com (V. Albiero), laurendeason@meta.com (L. Deason), ive@meta.com (I.-E. Veliche), christinelehane@meta.com (C. Lehane).

(Barrington et al., 2025; Kassis & Hengartner, 2023). Recent work has demonstrated that LLMs can generate sophisticated vishing transcripts that evade machine learning classifiers while preserving their deceptive content (Li et al., 2025), suggesting that defensive technologies may struggle to keep pace with offensive capabilities.

From an attacker's perspective, AI-powered vishing enables drastic cost reduction by eliminating the need for paid call center operators. It also provides significant scalability as AI agents can conduct thousands of concurrent calls, limited mainly by telephony and network capacities rather than human capacity. AI vishing also enables personalization at scale as LLMs can synthesize personal data from social media, data breaches, and public records into emotionally resonant narratives tailored to individual targets. This will have significant consequences to the economics of voice scams, which are already causing substantial harm. According to industry surveys, more than 56 million Americans were affected by scam calls in 2023, with collective losses exceeding $25 billion (LaMont, 2024).

The U.S. Federal Communications Commission explicitly ruled AI-generated voices in robocalls illegal under the Telephone Consumer Protection Act (FCC, 2024). These regulatory responses are important, but they are outpaced by technical change. Despite the attention, empirical evidence on how real people respond to AI-driven vishing remains sparse, as further shown in Section 2. To address this gap, we conducted the first large-scale, controlled evaluation of AI-powered voice phishing using a representative sample of 4,100 U.S. adults. Participants were randomly assigned to hear audio recordings or read transcripts of scam conversations generated by leading voice AI systems, including models from Meta (Llama Full-Duplex), OpenAI, Google (Gemini), Sesame, Play.AI, and ElevenLabs, alongside human baseline conditions. We complemented this quantitative assessment with 12 qualitative interviews to understand the perceptual and reasoning processes underlying susceptibility. Our paper makes four key contributions that span empirical novelty (new population-level evidence on susceptibility to AI-enabled vishing), analytical insight (the psychological mechanisms that predict compliance), and methodological novelty (a large-scale, nationally representative experiment combining multiple AI voice systems, human and text baselines, neutral controls, and qualitative interviews):

- We provide the first population-level estimates of susceptibility to AI-powered vishing, showing that compliance rates reached as high as 36.1% for emotionally personalized scams such as cloned relative-in-distress scenarios. Even when averaged across all scam categories, 16.5% of participants indicated that they would or might comply, an alarming level of susceptibility given the low cost and high scalability of AI-automated voice phishing. We use the term "compliance" to denote self-reported willingness to comply with the caller's request. We interpret this measure as an indicator of susceptibility to the scam rather than as observed behavioral compliance.
- We demonstrate that caller persuasiveness rather than human-likeness was the strongest predictor of compliance. This suggests that the psychological content and delivery of scam narratives matter more than achieving perfect vocal fidelity.
- We compare six leading AI voice systems against human baselines, finding that the Sesame model achieved ratings statistically indistinguishable from human voices across sentiment, persuasiveness, trustworthiness, and human-likeness, while other models scored significantly lower.
- We show that neither general AI familiarity nor voice assistant usage improves detection accuracy, indicating that prior exposure to AI does not meaningfully increase a user's protection against AI-powered deception.

Our findings suggest that the technical capability for scalable, automated vishing has arrived. The most advanced voice synthesis systems can now rival human callers in scam contexts, particularly when paired with emotionally charged scripts that exploit trust and urgency. Given the rapid pace of improvement in voice AI capabilities, these results likely represent a lower bound on future risk, with the vulnerabilities we document expected to grow as the technology matures. These results carry implications for consumer protection, platform governance, and AI model release policies that we discuss in subsequent sections.

## 2. Related work

**Phishing and Voice Phishing.** Phishing has long been studied as a socio-technical security problem that exploits human trust in addition to technical vulnerabilities (Dhamija et al., 2006). Early work established variation in susceptibility across users and contexts (Ribeiro et al., 2024; Sheng et al., 2010), motivating population-level measurement. Complementing the usable-security literature, a broad synthesis of fraud victimization research finds consistent psychological predictors of susceptibility (e.g., impulsivity and low self-control), while also emphasizing that risk factors vary by scam type and context rather than being explained by demographics alone (Dadà et al., 2025). Empirical evidence for voice phishing was provided by Tu et al. (2019), who conducted a large-scale telephone phishing experiment ($N \approx 3,000$), showing that users answer scam calls and disclose sensitive information. Analysis of 86 real-world vishing attacks reveals that social engineers most commonly exploit authority, social proof, and distraction as persuasion principles, with specific implementation patterns varying across attack types (Jones et al., 2021). Automated detection of such persuasion techniques in phishing emails has also been explored using transformer-based models (Jáñez-Martino et al., 2025). Figueiredo et al. (2024, 2025) demonstrate the technical feasibility of fully automated vishing systems in controlled settings. Complementary defensive work explores detection of voice phishing through system-level signals (Lee et al., 2025). On the technical detection side, hybrid deep learning frameworks have been proposed for automated phishing detection (Prasad et al., 2026). Qualitative interviews with deepfake fraud victims reveal that individuals associate AI-generated media with entertainment rather than threat (Zhang et al., 2025b). Our work complements this line of research by providing the first large-scale, controlled evaluation of human responses to AI-generated vishing, across multiple commercial and open-source voice models.

**Generative AI and Automated Social Engineering.** Advances in large language models have reduced the cost of personalization and enabled automation at scale (Carlini et al., 2025). Most LLMs can be exploited to bypass safety constraints and generate phishing messages that perform on par with those crafted by human experts (Gupta et al., 2023; Heiding et al., 2026, 2024). At the psychological level, Matz et al. (2024) demonstrate that generative models can produce messages tailored to individual traits, significantly increasing persuasive impact. In a comprehensive survey of vishing research, Triantafyllopoulos et al. (2025) note that most literature lacks controlled experiments measuring human susceptibility to modern AI-powered voice attacks.

**Voice Detection and Human Perception.** Human listeners have trouble consistently distinguishing AI-generated voices from their real counterparts, even after training (Barrington et al., 2025; Mai et al., 2023; Warren et al., 2024). In a gamified experiment with 472 participants, IT expertise provided no detection advantage, though native speakers outperformed non-native speakers (Müller et al., 2022). In a vishing-specific study, Bhatti et al. (2026) found that participants performed below chance at distinguishing AI-generated from human voices, relying on paralinguistic heuristics that modern synthesis systems readily replicate. Algorithmic detection models lag behind synthesis quality and generalize poorly across datasets (Zhang et al., 2025a), and adversaries can adapt content to evade classifiers (Li et al., 2025). Listeners form personality and trustworthiness judgments from voice samples within milliseconds based on acoustical clues (Belin et al., 2017; McAleer et al., 2014), and even synthetic voices can gain trust and persuade (Diel & MacDorman, 2024; Dubiel et al., 2020; Weber et al., 2020). Pias et al. (2024) demonstrated that varying the perceived tone, age, and gender of voice assistant speech directly influences purchase decisions. The existing work focuses on detecting AI-generated voices, rather than how detectability interacts with actual scam compliance. Our results show that detected AI voices can still be persuasive if they use compelling scripts, highlighting a gap between detectability and real-world protection.

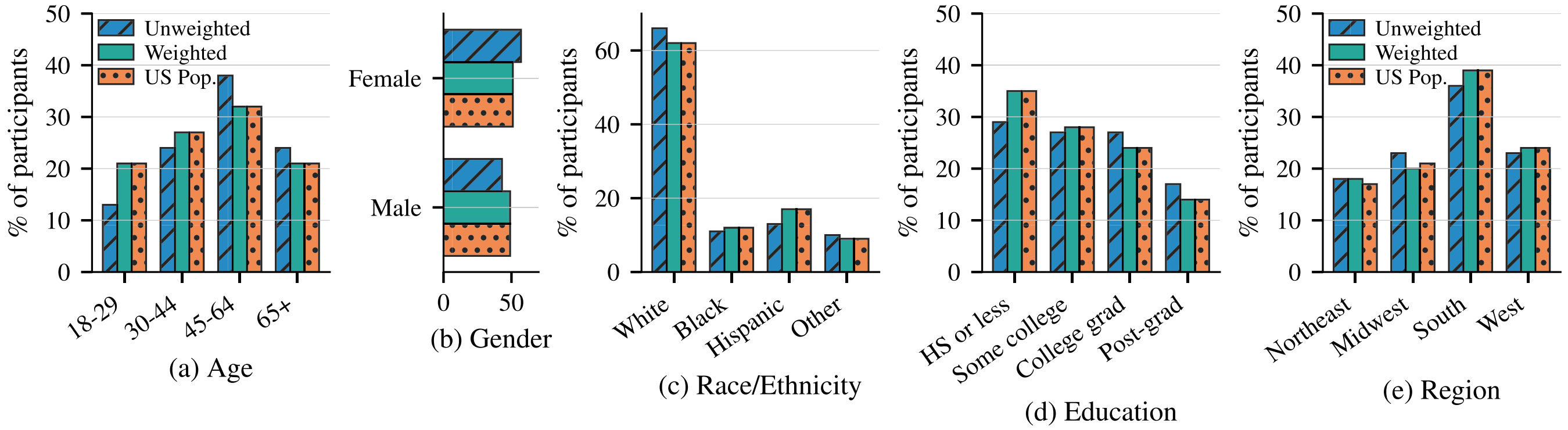

**Fig. 1.** Sample demographics. Unweighted sample (blue, hatched), weighted sample (teal), and US population benchmarks based on the 2023 Current Population Survey (orange, dotted). Survey weights were applied using age, gender, race/ethnicity, and education. The close alignment between weighted and US population distributions confirms successful post-stratification; the key comparison is between the unweighted sample and the target population. (a) Age. (b) Gender. (c) Race/ethnicity. (d) Education. (e) Region. Exact values are reported in our repository https://shorturl.at/gC8bH.

Armstrong et al. (2023) demonstrate that listeners default to assuming caller honesty during phone calls; highly sensitive requests reduce perceived honesty, but innocuous requests do not, and trust can partially recover even after suspicious triggers. Similarly, our qualitative interviews with participants show that they often attributed repetitive speech to a caller reading from a script, and unusual vocal characteristics such as rapid or uneven pacing to nervousness, rather than recognizing these as signs of AI generation. Prior work largely examines benign or commercial interactions rather than adversarial deception contexts. Our study extends these insights to malicious settings, showing how trust assumptions and persuasive framing shape vulnerability to AI-enabled voice manipulation.

## 3. Method

We conducted a large-scale survey to assess population-level susceptibility to AI-powered voice phishing and voice cloning attacks. The study employed a between-subjects design with 4,100 U.S. adults randomly assigned to one of 37 experimental conditions. We compared six AI voice models with human control voices across five scam scenarios. For each scam scenario, we also included a neutral condition: a matched, non-scam version of the same interaction that removed deceptive or manipulative elements while maintaining the same structure. This enabled a systematic comparison of AI and human persuasiveness at varying levels of emotional intensity and personalization. In each condition, participants were exposed to either an audio recording, a text transcript of a conversation generated using a specific AI voice model and scam scenario, or its respective human or neutral control. We complemented this quantitative experiment with 12 semi-structured qualitative interviews to capture participant perceptions of realism, persuasiveness, and scam detection.

### *3.1. Participants and recruitment*

We recruited 4661 participants through YouGov's opt-in panel (YouGov, 2025) between May 15–24, 2025. YouGov employs an opt-in online panel with diverse recruitment methods. Their panel members are recruited from a variety of sources, including through standard advertising and strategic partnerships with a broad range of websites. To ensure we accurately represent everyone in the country, we recruited participants from a wide range of backgrounds. YouGov offers surveys in different languages but for this survey, participants had to be able to understand English as the recordings were only made in English. All recruitment sources are monitored to ensure responsive and engaged participants. We partnered with YouGov to define non-sensical response patterns, and later remove them from the data. Non-sensical response examples include answering on extreme ends of the same scale, answering the same to reverse worded items, or answering too quickly to have paid attention. YouGov also identifies and removes duplicate participants, and uses a Defender Score to identify and remove responses from fraudulent respondents. YouGov used sample matching methodology to ensure representativeness of the internet, using U.S. adult population aged 18 or older. We constructed a target sample by grouping respondents based on the 2023 Current Population Survey and drawing weighted samples from each group (Cochran, 1977; Lohr, 2021). For each member of the target sample, we selected matching members from the opt-in panel based on age, gender, race/ethnicity, and years of education using propensity score matching (Rosenbaum & Rubin, 1983).

The 4661 respondents were then matched down to 4100 at the treatment level to produce balanced experimental groups (approximately $n = 111$ per condition). Weights were calculated using propensity scores. The matched cases and sampling frame were combined, and a logistic regression estimated inclusion probability based on age, gender, race/ethnicity, years of education, and region. Weights were then post-stratified on a four-way stratification of gender, age (4 categories), race (4 categories), and education (4 categories). The final weights ranged from 0.29 to 7.01, with a design effect of 1.21, yielding an effective sample size of 3388 and a margin of error of $\pm 1.85\%$ at the 95% confidence level. The final weighted sample of 4100 participants was representative of the U.S. internet-using adult population. The sample included 51.2% women, with a mean age of 45.3 years (SD = 16.8). Educational attainment varied: 28.3% high school or less, 31.4% some college, 25.1% bachelor's degree, and 15.2% graduate degree. The racial/ethnic distribution was: 62.4% White, 12.8% Black, 16.9% Hispanic, 5.3% Asian, and 2.6% Other. Fig. 1 presents the unweighted demographic composition. More details about the demographic sample composition are provided in our repository at https://shorturl.at/gC8bH.

### *3.2. Experimental design and scam scenarios*

**AI Voice Models.** We evaluated six AI voice systems: Llama Full-Duplex (experimental mode, Meta AI app, April 2025) (Veluri et al., 2024), OpenAI AVM (Sol voice) (Lin et al., 2025), Google Gemini (Ursa voice, mobile app only) (Team et al., 2023), Sesame (Sesame, 2025) (Maya voice, we used the web version but it is available as an open-source model), Play.AI (PlayAI, 2024) (Celeste voice), and ElevenLabs (ElevenLabs, 2024) (cloned voice). Our voice selection was guided by prior research demonstrating stronger user responses to young, female voices with neutral North American accents (Figueiredo et al., 2024; Pias et al., 2024). For the ElevenLabs model, we cloned a voice with an Irish female accent, and used a voice with a similar accent as the call recipient. With the exception of the grandma scam, this call recipient's voice is consistent across scenarios to avoid contamination by different voices.

The voice recordings were made via web versions, except for Gemini where voice mode was only available on mobile app. Because we did not have access to alter the system prompts directly in some platforms, we fed the prompts to competitor models through their standard interfaces and requested them to follow the role play instructions. For Llama FD, we got early access to the model launched in experimental mode in the Meta AI app on April 29, 2025. All recordings were made in April 2025.

**Scam Scenarios.** We designed five scam scenarios after consultations with subject matter experts from industry and academia, and inspired by literature such as Mitnick and Simon (2003), Vishwanath (2022). The experts provided a list of 26 representative examples of voice phishing scams currently deployed in the field, and we narrowed the list to viable scenarios that covered different types of phishing requests and could be feasibly evaluated within the study design. These scenarios varied by information type requested (credentials versus financial transactions) and caller-recipient relationship (known versus unknown). The scenarios were:

i. **MasterCard scam**: Credential phishing requesting credit card details, expiration date, and security code from an unknown caller impersonating MasterCard support.
ii. **Gmail scam**: Credential phishing requesting email login credentials and 2FA code from an unknown caller impersonating Google support.
iii. **Donation scam**: Financial transaction request for a cash transfer from an unknown caller soliciting donations for a charitable cause.
iv. **Police-Grandma scam**: Financial transaction request where an unknown caller impersonating police claims the recipient's relative is in trouble and needs bail money.
v. **Sister-in-distress scam (ElevenLabs)**: Financial transaction request using a cloned voice where a supposedly known caller (the recipient's sister) claims to be in an emergency and needs immediate financial help.

**Conversation Design.** All interactions followed standardized 10-turn conversations that ended ambiguously, with the call recipient neither agreeing nor refusing the request. This design ensured comparability across conditions while providing sufficient conversational depth for participant evaluation. We used identical system prompts across AI models for each scenario and scripted responses for the human call recipient. The prompts instructed models to adopt personas of legitimate service representatives, create urgency, use guilt-based persuasion, and persist in gathering sensitive information. The AI-generated responses naturally varied across interactions, but the human call recipient followed a structured response script designed around consistent conversational intents and decision points rather than exact wording, allowing for minor variation while preserving comparability across conditions.

To simulate malicious actor deployment, we edited recordings in Adobe Audition to remove safety mitigations (refusals, disclaimers, role-play deviations) and UX features that clearly signaled AI generation (turn-taking beeps). This represents the cleanest possible version of each interaction. If scammers deployed these models without such editing, their likelihood of success would be lower due to these built-in safety features.

**Comparison Conditions.** We implemented several control conditions to isolate specific factors contributing to scam effectiveness. Each AI scam condition included a corresponding neutral control scenario using the same voices but benign content. Our experimental design allowed us to disentangle the components contributing to persuasiveness and human-likeness through four types of comparisons:

1. *Voice and content source*: We compared conditions where (a) both voice and content were AI-generated, (b) voice was human but content was AI-generated, (c) both voice and content were human-generated, and (d) voice was AI-generated but content was human-generated. This allowed us to isolate the independent effects of voice quality and script content.
2. *Scam versus neutral content*: By comparing scam scenarios to neutral controls using identical voices, we assessed content-driven effects independent of voice quality.
3. *Voice versus text modality*: We included text transcript conditions presenting the same conversational content without audio, allowing us to identify whether voice has an impact on persuasiveness or human-likeness beyond the content itself.
4. *Relational closeness*: The donation, police-grandma, and cloned sister scenarios all incorporated a personal appeal (someone in need) and requested a cash transfer, but differed in the degree of relational closeness between the call recipient and the caller or beneficiary. Comparing these conditions allowed us to estimate the differential effect of familiarity on compliance.

Together, these four comparisons constitute an ablation-style decomposition of scam effectiveness. Each comparison isolates the independent contribution of a single factor (voice quality, script content, presentation modality, or relational closeness) against human-voice, neutral-content, and text-transcript baselines. In addition, we include a recording of a real conversation between human participants reenacting a scam scenario, serving as an ecological-validity benchmark to assess how closely the experimental conditions approximate real-world scams.

Human voice baseline conditions were created with an internal volunteer whose voice is audibly comparable to the main AI voices (young, female, neutral North American accent). We also included a recording of an actual scam call that targeted members of the public in North America to serve as an ecological validity check.

**Limitations of Experimental Design.** Our design prioritized ecological validity over perfect experimental controls. For instance, the donation scenario is not an ideal control for the sister-in-distress scenario; ideally, we would have used a cloned voice making a donation request. However, we selected the sister-in-distress scenario because it reflects current scam patterns more accurately. Additionally, the cloned voice in the sister scenario used an Irish accent rather than the North American accent of other conditions, which may affect comparability. All recordings maintained consistent nonprofessional quality across conditions to reflect realistic deployment scenarios. More information about the study's validity is presented in Section 6.2.

### 3.3. Survey instrumentation

Each participant evaluated one randomly assigned audio recording or transcript. The survey assessed five dimensions using validated scales where available and newly developed measures where necessary. In particular, it combined two single-item measures (*sentiment* and *persuasiveness*), two multi-item scales (*trustworthiness* and *human-likeness*), and one behavioral outcome (*willingness to comply*). Unless otherwise noted, all items were assessed using five-point Likert-type scales (Likert, 1932). These measures were selected to capture both participants' perception of the caller and their susceptibility to the scam request. Sentiment, trustworthiness, persuasiveness, and human-likeness assess how participants evaluate the caller and the interaction, while compliance reflects self-reported willingness to comply with the scam request. Together, these metrics provide a holistic assessment of AI-enabled vishing effectiveness.

**Caller sentiment**, that is, the likeability or initial impression, was measured with a single item adapted from prior AI voice phishing research (Figueiredo et al., 2024): *"How would you describe your initial impression of the caller?"* rated on a 5-point scale (1 = Very negative, 5 = Very positive).

**Persuasiveness** was assessed with a single item: *“Overall, how convincing did you find the caller during the conversation?”* rated on a 5-point scale (1 = Not at all convincing, 5 = Very convincing). Based on our pilot qualitative sessions and internal survey expert feedback, we identified that the concept “convincing” was most readily used and easy to understand. Therefore, we used a single-item focused on how convincing participants found the caller.

**Trustworthiness** was measured using a newly developed 9-item Caller Trustworthiness Scale. No validated trustworthiness measure appropriate for caller assessment existed in prior literature. The scale was developed by reviewing existing trust literature and incorporating concepts that emerged during pilot qualitative sessions. Participants rated the extent to which the caller demonstrated expertise, credibility, empathy, understanding, friendliness, compelling explanations, trustworthiness (direct item), confidence, importance/urgency, clarity/coherence on 5-point scales (1 = Not at all, 5 = Very much). We used exploratory factor analysis to examine the scale’s underlying structure. The Kaiser-Meyer-Olkin (KMO) measure verified sampling adequacy (KMO = 0.921), which is considered excellent (Kaiser, 1974), and Bartlett’s test of sphericity indicated sufficient correlations for factor analysis ($\chi^2(36) = 25,093.79$, $p < 0.001$) (Bartlett, 1954). One item (“importance/urgency”) was removed due to low extraction communality (.088). A single-factor solution was retained based on eigenvalues greater than 1 and scree plot examination, accounting for 58.66% of total variance. Factor loadings ranged from 0.67 to 0.83. We assessed the scale’s reliability using Cronbach’s alpha (Cronbach, 1951), which measures the extent to which scale items consistently measure the same underlying construct. The final 9-item scale (range: 9–45) demonstrated excellent internal consistency ($\alpha = 0.927$), well above the conventional threshold of 0.70 for acceptable reliability, supporting its use as a unidimensional measure of perceived caller trustworthiness.

**Human-likeness** was measured using the Partner Modelling Questionnaire (Doyle et al., 2025), validated specifically for voice-enabled AI agents. Participants rated bipolar adjective pairs on 5-point scales: from warm to cold, personal to generic, empathetic to apathetic, social to transactional, life-like to tool-like, and human-like to machine-like. Scores were summed to create a composite (range: 6–30).

**Compliance** was assessed with a behavioral intention question: *“If you were on the receiving end of this call, would you agree to the caller’s request?”* with response options Yes, No, or Unsure. We report the percentage answering Yes or Unsure as the compliance rate, acknowledging that subjective ratings may overestimate actual compliance behavior.

To complement our quantitative findings with deeper insights into participant reasoning and perceptions, we conducted 12 semi-structured interviews with U.S. adults recruited through AnswerLab. Full details of the interview methodology, sample composition, protocol, and data collection procedures are provided in Appendix D.

We note that an optimal design would have involved a representative sample of the U.S.-based, English-speaking population interacting directly with the helpful-only AI models or human controls via Twilio[1] (a cloud communication platform), followed by an online survey to capture their responses. For the voice clone condition, this would include that a known person to each participant would have their voice cloned and would also agree to participate in the study. We instead chose our final design to allow for a scalable evaluation of the persuasiveness and human-likeness of each model. Rather than having participants interact directly with the AI models – which could risk eliciting sensitive information like credit card details – we instead asked them to rate audio recordings or transcripts of potential scam calls, consistent with the boundaries set by our ethical review.

### 3.4. Statistical analysis

Data were weighted using propensity scores as described in Section 3.1 to ensure representativeness of the U.S. internet-using adult population. Statistical significance was set at $\alpha = 0.05$ for all tests. We report significance tests and effect sizes throughout the paper, and, where applicable, 95% confidence intervals, so that both statistical and practical significance can be assessed. Effect sizes are reported alongside $p$-values to facilitate interpretation of practical significance.

We assessed internal consistency for multi-item scales using Cronbach’s alpha (Cronbach, 1951), with values of $\alpha = 0.92$ for the 9-item Trustworthiness scale and $\alpha = 0.93$ for the 6-item Human-likeness scale (Partner Modelling Questionnaire), both well above the conventional threshold of 0.70 for acceptable reliability (Nunnally, 1978). To compare continuous outcomes (caller sentiment, persuasiveness, trustworthiness, and human-likeness) across AI models within each scam scenario, we conducted one-way analysis of variance (ANOVA). Prior to analysis, we assessed homogeneity of variance using Levene’s test (Levene, 1960). Three of four outcomes showed significant heterogeneity of variance: sentiment ($W = 3.24$, $p < 0.001$), persuasiveness ($W = 2.13$, $p < 0.001$), and human-likeness ($W = 2.31$, $p < 0.001$); trustworthiness did not violate the assumption ($W = 1.29$, $p = 0.15$). Consequently, we employed Welch’s ANOVA (Welch, 1951), which does not assume equal variances and provides valid inference under heterogeneity. For statistically significant omnibus tests, we conducted Bonferroni-corrected post-hoc tests to control for Type I error inflation (Maxwell et al., 2024). Effect sizes for ANOVAs are reported as eta-squared, with interpretation following conventional guidelines: small ($\eta^2 = 0.01$), medium ($\eta^2 = 0.06$), and large ($\eta^2 = 0.14$).

When comparing each AI model to the human voice control within a scenario, we employed Dunnett’s $t$-tests (Dunnett, 1955), which treat the human condition as the reference group and provide greater statistical power than standard pairwise comparisons for multiple-to-one contrasts (Hsu, 1996). For specific planned comparisons between two conditions (e.g., AI voice versus text transcript, scam versus neutral content), we conducted independent samples $t$-tests. Where Levene’s test indicated unequal variances, we applied Welch’s $t$-test (Welch, 1947). We report Cohen’s $d$ as the measure of effect size for pairwise comparisons, with interpretation following conventional guidelines: small ($d = 0.2$), medium ($d = 0.5$), and large ($d = 0.8$) (Cohen, 1988).

For the primary outcome of willingness to comply (Yes/No/Unsure), we employed chi-square tests of independence to assess associations between experimental conditions and compliance responses. We report Cramér’s $V$ as the effect size measure (Cramér, 1946), with the standard interpretation small ($V = 0.1$), medium ($V = 0.3$) and large ($V = 0.5$).

To identify factors associated with compliance while controlling for potential confounds, we conducted weighted binary logistic regression with compliance (Yes or Unsure versus No) as the dependent variable. Predictor variables included experimental condition, demographic characteristics (age, gender, education, race/ethnicity), AI familiarity, and psychological measures (trustworthiness, persuasiveness, human-likeness, sentiment). Multicollinearity was examined using variance inflation factors (VIF), with all values below 5 (range: 1.03–2.65), indicating acceptable levels (Hair et al., 2019). Model fit was assessed using Nagelkerke’s $R^2$ (Nagelkerke, 1991) ($R^2 = 0.35$) and classification accuracy (83.7% correctly classified). We report odds ratios (OR) with 95% confidence intervals, where OR $> 1$ indicates increased odds of compliance and OR $< 1$ indicates decreased odds.

**Qualitative Analysis.** We conducted 12 semi-structured interviews to complement the quantitative findings with deeper insights into participant reasoning and perceptions. Sample size was informed by practical constraints and established guidelines suggesting that thematic saturation for focused research questions often occurs within 10–15 interviews (Guest et al., 2006). We acknowledge that our deliberately diverse sample may limit saturation across demographic subgroups; however, the qualitative component was intended to illuminate mechanisms underlying the quantitative findings rather than to achieve standalone generalizability. Interviews were audio-recorded, transcribed verbatim, and analyzed using thematic analysis following Braun and Clarke's six-phase framework (Braun & Clarke, 2006). A single researcher conducted all coding, precluding inter-rater reliability assessment. Rigor was maintained through iterative review and by confirming themes recurred across the dataset.

[1] https://www.twilio.com/.

## 4. Results

This section presents findings from our large-scale evaluation of AI-powered voice phishing across multiple dimensions. We organize our results to address our primary research questions systematically, examining (i) how AI models perform in neutral contexts, (ii) how scam content affects perception, and (iii) what factors drive susceptibility to AI-powered attacks. We organize our analysis to trace the progression from baseline AI capabilities through the transformation imposed by malicious intent, examining compliance patterns, conducting systematic comparisons across modalities and actors, and finally assessing detectability, thereby providing a comprehensive picture of the AI-powered vishing threat landscape.

First, we establish baseline performance differences between AI models in neutral (non-scam) contexts to understand their inherent qualities independently of malicious use (Section 4.1). We then examine how the introduction of scam content affects the user perception of these models (Section 4.2), before analyzing actual compliance rates and factors that predict susceptibility to AI-powered scams (Sections 4.3 and 4.7). Subsequently, we compare AI model performance against human voice controls (Section 4.5) and investigate the differential effects of voice versus text modality (Section 4.6). We also analyze how different scam types, from generic account support to personalized emotional appeals, impact effectiveness (Sections 4.3 and 4.4). Finally, we examine users' ability to detect AI-generated voices (Section 4.8), providing insight into the current detectability of these systems. Throughout this section, all reported statistics are weighted to represent the U.S. internet-using adult population, and statistical significance is set at $\alpha = 0.05$.

### 4.1. Establishing a baseline: AI model performance in neutral scenarios

To establish baseline differences between AI models independent of scam context, we first compared participant responses to neutral (non-scam) scenarios where a school administrator notifies a parent about school closure dates.

In neutral scenarios, all AI-generated callers displayed comparable sentiment (mean ratings 3.85–4.09 on a 5-point scale), with no significant differences across models. However, clear differences appeared in perceived human-likeness: Sesame was rated as significantly more human-like than Llama FD ($p = 0.035$), OpenAI AVM ($p = 0.039$), and Gemini ($p = 0.005$), but did not differ from Play.AI or ElevenLabs.

When compared against an authentic human voice, Sesame was the only model that achieved statistical parity on human-likeness ($p = 0.135$), while Llama FD, Play.AI, OpenAI AVM, and Gemini all scored significantly lower (all $p < 0.001$). The ElevenLabs cloned voice also matched human-level performance on both sentiment ($p = 0.051$) and human-likeness ($p = 0.661$), suggesting that high-quality voice cloning can achieve human-level naturalness in neutral contexts. Complete pairwise comparisons and visualizations are provided in Appendix E.1.

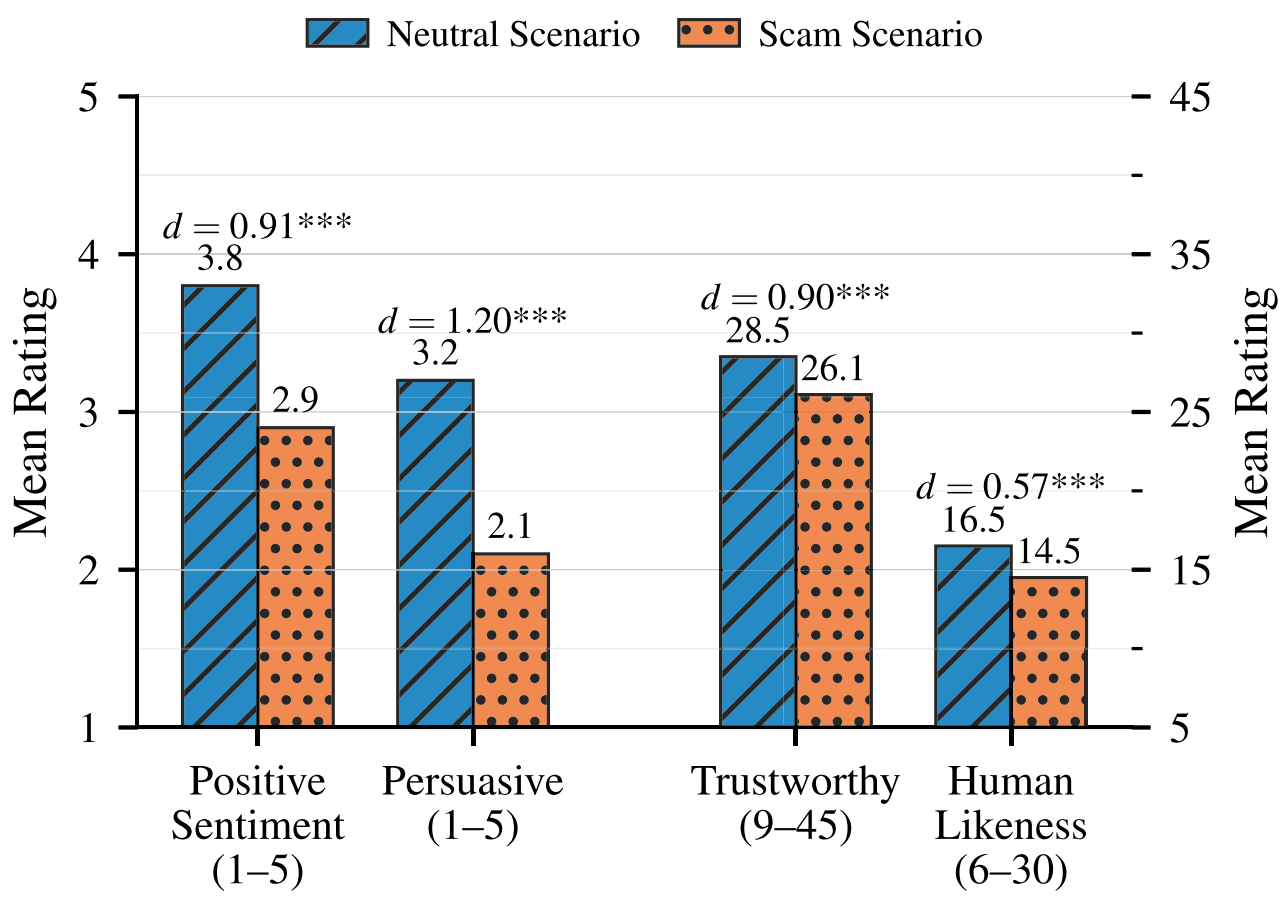

**Fig. 2.** Impact of scam context on AI model perception. Mean ratings across four key perception dimensions comparing neutral (non-scam) and scam scenarios for combined AI models (Llama FD, Sesame, Play.AI, OpenAI AVM, and Gemini). Effect sizes (Cohen's $d$) indicate large negative effects of scam context across all dimensions. All differences significant at $p < 0.001$. Caller Sentiment and Caller Persuasiveness are measured on 1–5 scales; Caller Trustworthiness on a 9–45 scale; Caller Human-Likeness on a 6–30 scale.

### 4.2. The scam context effect: how malicious intent transforms perception

*Introducing scam content dramatically degraded all model perceptions.* The transition from neutral to scam scenarios produced negative effects across all measured dimensions. Fig. 2 illustrates the impact of the scam context on participants' perception in all measured dimensions. The introduction of scam content resulted in significantly lower ratings compared to neutral scenarios. The strongest impact was on persuasiveness (Cohen's $d = 1.20$), followed by caller sentiment ($d = 0.91$) and trustworthiness ($d = 0.90$). Scam content even reduced perceived human-likeness ($d = 0.57$), suggesting that suspicion affects perceptual judgments.

*This pattern held consistently across both voice and text modalities.* Text transcripts showed the same moderate-to-large negative effects ($p < .05$), confirming that the scam scenario itself, rather than voice-specific artifacts, drives increased suspicion. However, as we demonstrate below in Section 4.6, the quality of the AI voice determines whether this suspicion translates into actual protection.

### 4.3. Compliance rates: when AI-powered scams succeed

Overall compliance with AI-powered scam requests averaged 16.5% (yes/unsure), with substantial variation by scam type, message framing, and voice model.

*Personalized, emotionally charged scams dramatically outperformed generic account support scams.* Relative to the MasterCard baseline, participants were significantly more likely to comply with scams invoking personal or emotional appeals as shown in Fig. 3. Compliance odds were approximately 3 times higher for donation requests (OR = 3.0, $p < 0.001$), 3.1 times higher for the police-grandma scenario (OR = 3.1, $p < 0.001$), and 5.33 times higher for the cloned-sister-in-distress scenario (OR = 5.33, $p < 0.001$). In contrast, the Gmail support scam did not differ significantly from MasterCard ($p > 0.05$), indicating that generic account recovery messages elicit consistently low compliance regardless of brand.

Among the unknown caller scenarios (MasterCard, Gmail, and Donation), which all used the same five AI voice models, the donation scam achieved the highest compliance rate, suggesting that scam framing alone substantially affects compliance. Scams incorporating personal appeals, whether through emotional connection to a known person (grandma scenario) or through voice cloning (sister scenario), demonstrated even higher effectiveness. However, these scenarios used different voice models (Sesame and ElevenLabs, respectively), so their elevated compliance reflects both personalization and voice model differences.

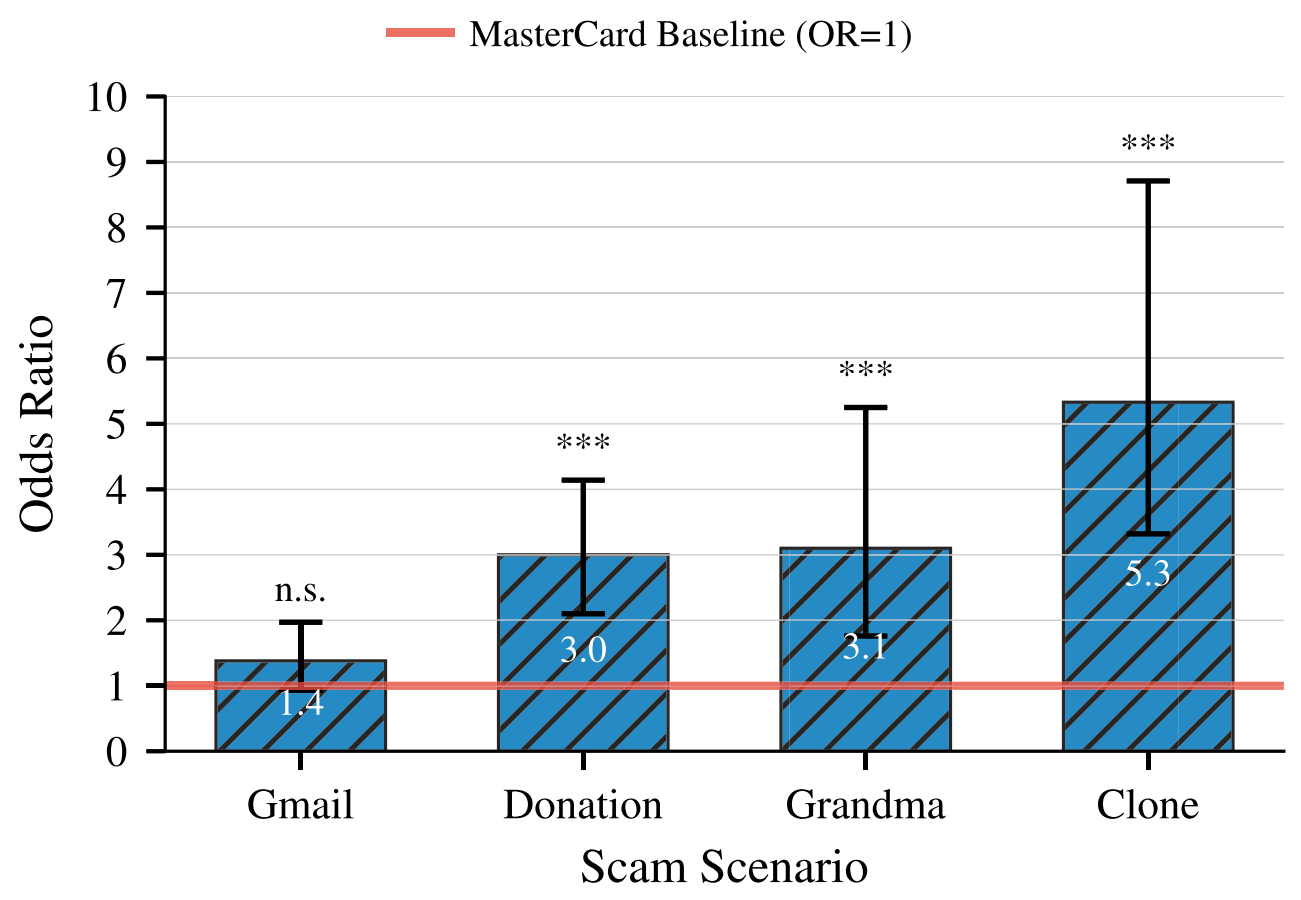

**Fig. 3.** Willingness to comply with scam requests by scenario type. Odds ratios (Exp(B)) show compliance likelihood (unsure/yes responses) relative to the MasterCard scam baseline (red dashed line at 1.0). Error bars represent 95% confidence intervals. $^{**}p < 0.01$; $^{***}p < 0.001$; n.s. = not significant. Personal appeal scams (Donation, Grandma, Clone) produced 3–5× higher compliance odds than generic account support scams.

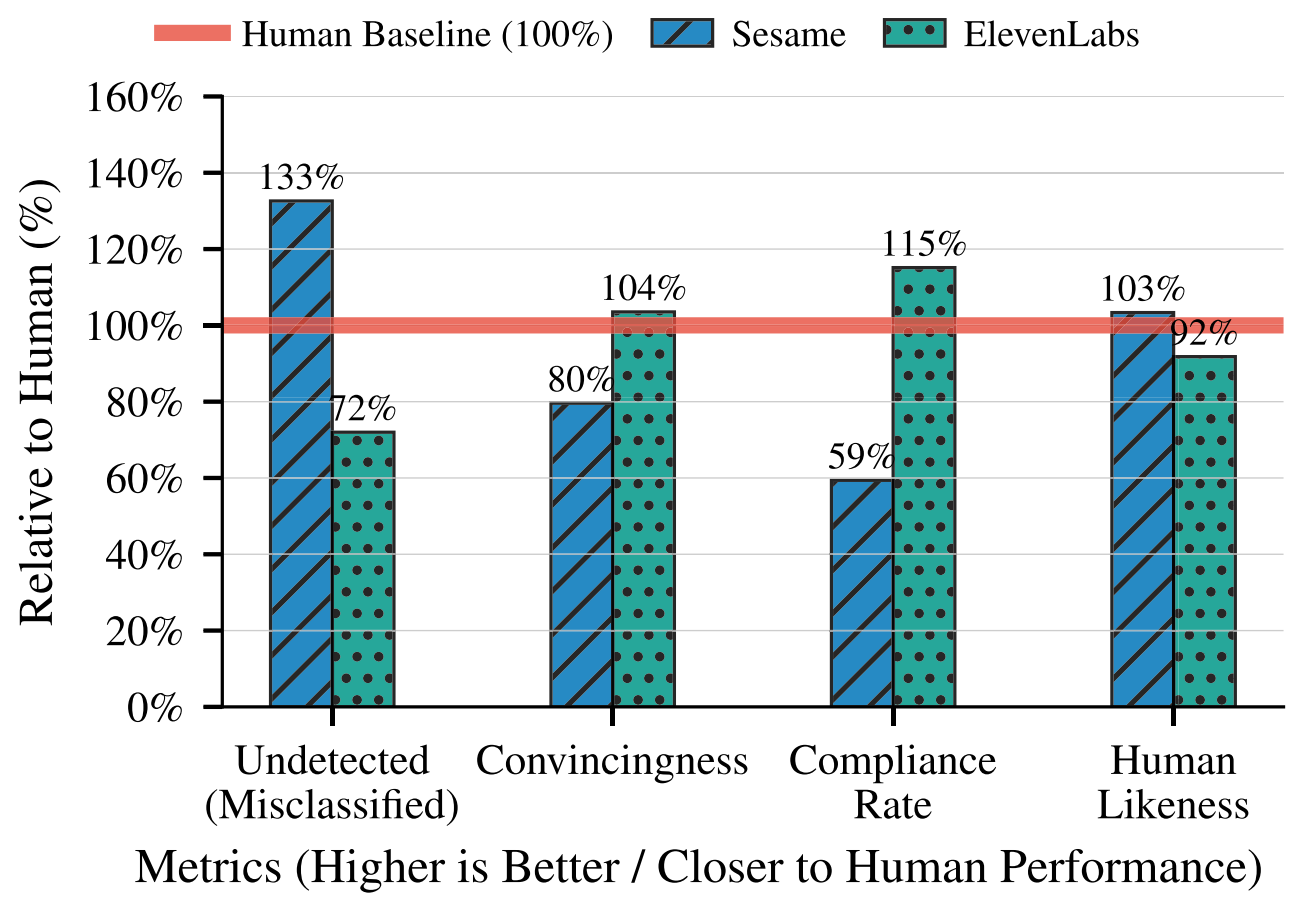

**Fig. 4.** AI voice performance relative to human baseline across two family emergency vishing scenarios with two models. All metrics normalized to human performance.

*The most effective scam, the ElevenLabs cloned-voice sister-in-distress, achieved the highest compliance rate (36.1%).* This represents more than a five-fold increase over the least effective baseline scenario, underscoring the persuasive power of emotionally evocative, relationship-based messages. The human-voiced grandma and donation scams also showed elevated compliance (24.1% and 32.6%, respectively; see Table E.4), suggesting that appeals invoking empathy or personal connection reliably increase susceptibility across multiple implementations.

### 4.4. AI system risk comparisons

Having established that scam context affects perception and that scam type drives compliance, we now examine whether specific AI models are more persuasive than others. We analyze model performance separately for unknown caller scenarios (generic institutional scams) and family emergency scenarios (personalized emotional appeals), as these contexts present fundamentally different attack surfaces. We use Llama FD as the reference model because it consistently represents a lower bound across key outcome dimensions, including sentiment, persuasiveness, trustworthiness, and human-likeness. Using a lower-quality AI voice rather than a human baseline also allows us to isolate how improvements in voice quality affect scam effectiveness across AI systems.

#### 4.4.1. Unknown caller scams

*Sesame outperformed competing AI models.* One-way ANOVAs revealed significant differences among AI models across all four dimensions: sentiment ($\eta^2 = 0.22$), persuasiveness ($\eta^2 = 0.24$), trustworthiness ($\eta^2 = 0.23$), and human-likeness ($\eta^2 = 0.16$). In scam scenarios involving unknown callers (MasterCard, Gmail, donation), Sesame yielded significantly higher ratings than Llama FD on caller sentiment ($p = 0.005$), persuasiveness ($p = 0.023$), trustworthiness ($p < 0.001$), and human-likeness ($p < 0.001$). Play.AI also performed significantly better than Llama FD on trustworthiness and human-likeness (both $p < 0.05$). Complete pairwise comparisons are provided in Table E.3 in Appendix E.2.

*However, model differences did not translate into significantly different compliance rates.* Despite Sesame's perceptual advantages, compliance rates did not differ significantly across models ($\chi^2(4) = 6.4, p = 0.17$). This suggests that while users can perceive quality differences between AI voices, these differences may not sufficiently alter behavior in generic scam contexts. In Section 4.5 below, we will see that the relationship between voice quality and compliance becomes more pronounced in personalized scenarios.

#### 4.4.2. Family emergency scams

*In high-stakes emotional scenarios, AI models approached or exceeded human baseline performance* as shown in Fig. 4. For the grandma scam, Sesame achieved 103% of human persuasiveness, 98% of human trustworthiness, and 100% of human human-likeness ratings. The ElevenLabs cloned voice showed similar performance in the sister-in-distress scenario, reaching 92% of human persuasiveness and 95% of human human-likeness. Notably, in these high-stakes emotional contexts, both Play.AI (in the donation scenario) and ElevenLabs (in the relative-in-distress scenario, involving a generic familial emergency, no voice cloning) were rated as more persuasive, trustworthy, and better liked than their human counterparts.

However, in the sister-in-distress scam–where ElevenLabs was used to clone the voice of a known individual–the AI-generated voice received lower persuasiveness and human-likeness ratings than the authentic human voice it impersonated. This divergence may reflect heightened scrutiny in identity sensitive contexts or limitations in conveying genuine emotional distress. Nevertheless, willingness to comply did not significantly differ between the AI and human voice conditions ($p > 0.05$), indicating that even imperfect voice cloning can remain effective in emotionally charged scams.

These results demonstrate that emotionally urgent, relationship-based scams amplify the effectiveness of high-quality AI voices, bringing them to parity with, or even beyond, human callers in several key dimensions, even when perceptual quality shows some degradation under scrutiny.

Table E.4 summarizes compliance rates across all AI voice and control conditions. While model-level differences were minor, scenarios combining human-like voices with emotional context (e.g., cloned or familial voices) elicited notably higher compliance than neutral or technical-sounding messages.

#### 4.4.3. Scenario comparisons

*Across all scam scenarios, scam type strongly influenced outcomes more than model choice.* Personal appeal scams (donation, police-grandma, relative-in-distress) consistently drove higher persuasion and compliance than account support scams (Gmail, MasterCard), with donation scams achieving compliance rates above 20% for several AI voices while account support scams rarely exceeded 15% (see Table E.4 in Appendix E.3 for complete condition-level rates). These findings highlight that social engineering content and emotional framing may matter

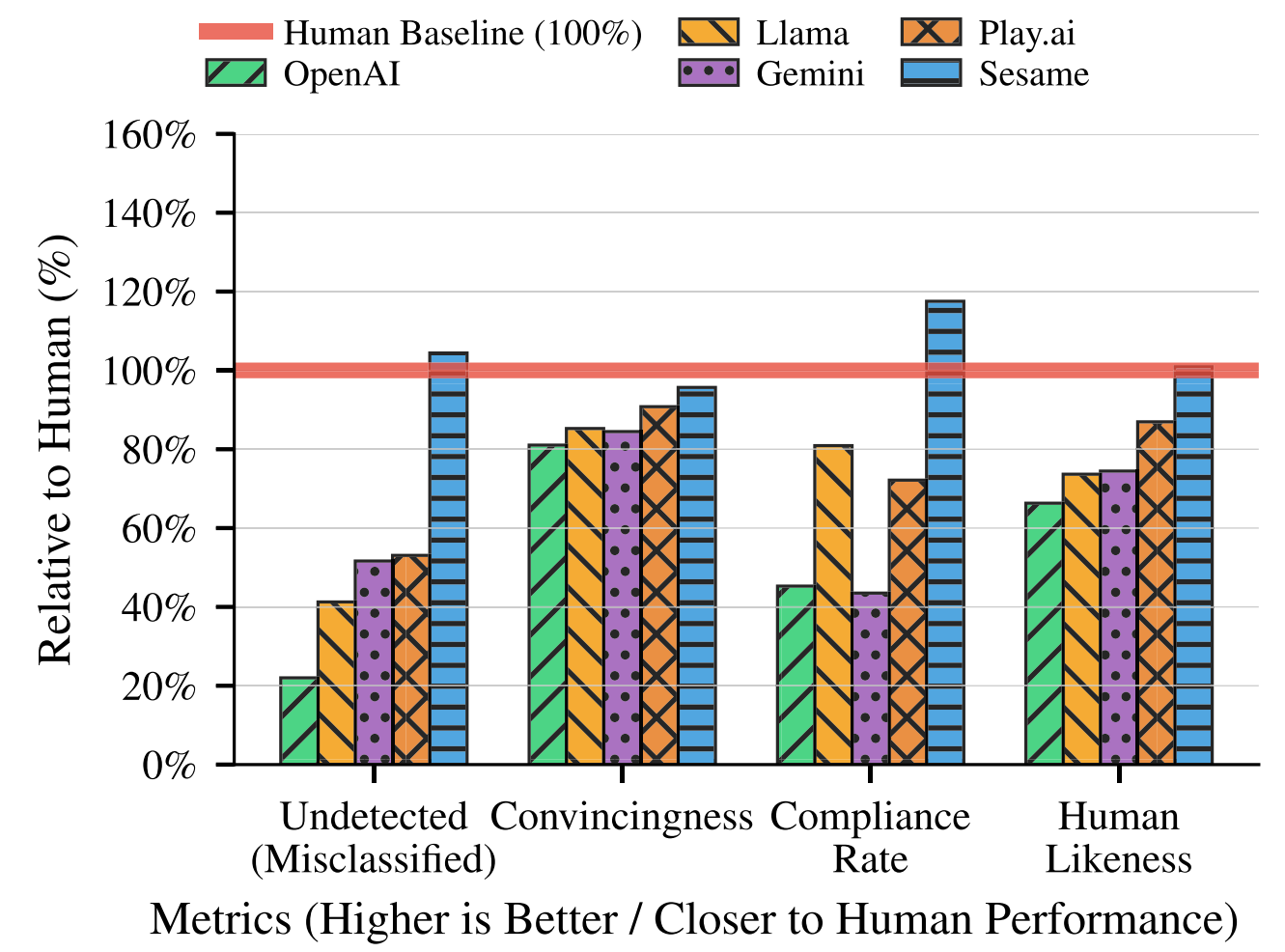

**Fig. 5.** AI voice performance relative to human baseline across three vishing scenarios with 5 models (MasterCard fraud, Gmail compromise, charity donation; n = 2,003). All metrics normalized to human performance (red line = 100%).

**Table 1**
AI models vs. human voice in scam scenarios.

| Model | Sentiment | Persuasive | Trustworthy | Human-Like |
|---|---|---|---|---|
| Llama FD | −0.344* | −0.369* | −4.556* | −3.749* |
| | (.003) | (<.001) | (<.001) | (<.001) |
| Sesame | −0.022 | −0.108 | −0.983 | −0.115 |
| | (1.000) | (.685) | (.433) | (.999) |
| Play.AI | −0.198 | −0.230 | −2.622* | −1.639* |
| | (.169) | (.065) | (<.001) | (.002) |
| OpenAI AVM | −0.333* | −0.474* | −3.814* | −3.873* |
| | (.003) | (<.001) | (<.001) | (<.001) |
| Gemini | −0.306* | −0.388* | −4.247* | −3.542* |
| | (.010) | (<.001) | (<.001) | (<.001) |

Values represent mean differences between each AI model and the human voice baseline. Negative values indicate lower ratings than the human voice. Numbers in parentheses are p-values from Dunnett's tests. ∗ indicate statistical significance ($p < 0.05$).

more than technical voice quality alone, and further demonstrate the importance of context-based spam filters, as proposed by Heiding et al. (2026). Nevertheless, the relative performance of AI models was stable across scam types, with Sesame and Play.AI often approaching human baselines.

### 4.5. AI versus human voices: closing the authenticity gap

*Sesame alone achieved parity with human voices across all scam scenarios.* As illustrated in Fig. 5, when comparing AI-powered scams to authentic human-voice scams, Sesame was the only model that performed comparably to human voices across sentiment, persuasiveness, trustworthiness, and human-likeness (Table 1). Llama FD, OpenAI AVM, and Gemini all scored significantly lower than human voices on sentiment and persuasiveness (all $p < 0.05$), while also underperforming on trustworthiness and human-likeness. Notably, some AI models (Play.AI and ElevenLabs) were more persuasive, trustworthy, and better liked than their human counterparts for the donation and relative-in-distress scam scenarios.

*Human voices still showed higher overall compliance.* There was a significant association between voice type (AI vs. human) and willingness to comply ($\chi^2(2) = 8.07, p = 0.018$, Cramér's $V = 0.09$), with human voice scams achieving 21.4% compliance compared to 15.4% for AI voice scams. However, this advantage varied by scenario and model quality, with Sesame and ElevenLabs approaching or matching human performance in specific contexts.

**Table 2**
Intercorrelations between primary variables in scam conditions.

| Variable | 1 | 2 | 3 | 4 |
|---|---|---|---|---|
| 1. Sentiment | – | | | |
| 2. Persuasiveness | 0.47** | – | | |
| 3. Trustworthiness | 0.52** | 0.66** | – | |
| 4. Human-Likeness | 0.43** | 0.61** | 0.70** | – |
| *M* | 2.87 | 2.12 | 26.05 | 14.54 |
| *SD* | 1.30 | 1.21 | 8.76 | 6.28 |

*Note:* $^{**}p < 0.001$. $N$ = 1,994 (combined AI voice scam conditions). All results weighted.

### 4.6. Voice quality moderates the effectiveness of audio versus text

*Voice quality determines whether audio enhances or diminishes scam effectiveness.* For lower-quality voice such as Llama FD and OpenAI AVM, text transcripts showed marginally higher compliance than voice calls (e.g., Llama FD: $\chi^2(2) = 6.06, p = 0.048$) and were rated as more human-like (Cohen's $d = 0.42, p < 0.001$). In contrast, for the higher-quality Sesame voice delivering the grandma scam, the voice condition was rated as significantly more persuasive ($p = 0.009, d = 0.38$), trustworthy ($p = 0.047, d = 0.28$), and human-like ($p = 0.010, d = 0.36$) compared to text alone.

These findings resolve an apparent paradox: while participants generally find it harder to detect AI in text versus voice, see Section 4.8, high-quality AI voices can nevertheless increase scam success relative to text by adding persuasive vocal cues. Conversely, low-quality voices may introduce suspicious artifacts that text avoids, potentially alerting victims. This suggests that voice deployment is not universally advantageous. Indeed, attackers must achieve sufficient quality for audio to enhance rather than undermine their deception. The greater variability in outcomes across scenarios in voice conditions compared to text conditions suggests that these differences stem primarily from voice characteristics rather than script content. This finding underscores the importance of voice quality as an independent factor in scam effectiveness.

### 4.7. What predicts compliance? Persuasiveness over human-likeness

All four measured variables, namely, caller sentiment, persuasiveness, trustworthiness, and human-likeness, were significantly and positively intercorrelated (Table 2), with correlations ranging from $r = 0.43$ (sentiment and human-likeness) to $r = 0.70$ (trustworthiness and human-likeness; all $p < 0.001$). Despite these correlations, binary logistic regression analysis revealed distinct patterns in their predictive power.

*Persuasiveness emerged as the strongest predictor of scam susceptibility.* Binary logistic regression analysis, measuring compliance (Yes or Unsure = 1) versus non-compliance (No = 0), identified caller persuasiveness as the strongest predictor of willingness to comply with scam requests (OR = 2.58, 95% CI [2.14, 3.10], $p < 0.001$), followed by caller sentiment (OR = 1.64, 95% CI [1.36, 1.97], $p < 0.001$) and caller trustworthiness (OR = 1.48, 95% CI [1.28, 1.71], $p < 0.001$).

*Contrary to expectations, human-likeness did not independently predict compliance.* When controlling for persuasiveness, sentiment, and trustworthiness, caller human-likeness failed to reach significance (OR = 1.15, 95% CI [0.92, 1.44], $p = 0.222$). While human-likeness correlated significantly with other variables (r = 0.43 to 0.70, all $p < 0.001$), it did not uniquely contribute to compliance beyond what was captured by psychological evaluations. This suggests that the content and delivery of the scam message matter more than achieving perfect vocal fidelity. We note that the intercorrelations among perceptual measures mean that this analysis identifies unique predictive contributions after

**Table 3**
Accuracy in identifying AI vs. human callers.

| Condition | N | Correct (%) | Incorrect (%) | Accuracy |
|---|---|---|---|---|
| *Text Transcripts* | | | | |
| AI-generated text | 693 | 53.0 | 47.0 | 53.0 |
| *Voice Recordings* | | | | |
| AI-generated voice (all) | 1994 | 70.3 | 29.7 | 70.3 |
| Human voice (controls) | 680 | 32.6 | 67.4 | 32.6 |
| *By AI Model (Voice, Scam Scenarios)* | | | | |
| Llama FD | 320 | 88.6 | 11.4 | 88.6 |
| OpenAI AVM | 351 | 93.7 | 6.3 | 93.7 |
| Gemini | 323 | 88.4 | 11.6 | 88.4 |
| Sesame | 343 | 69.1 | 30.9 | 69.1 |
| Play.AI | 342 | 85.6 | 14.4 | 85.6 |
| Sesame (Grandma) | 113 | 63.0 | 37.0 | 63.0 |
| ElevenLabs Clone | 107 | 75.8 | 24.2 | 75.8 |

*Note:* For human voices, "correct" = identified as human. For AI voices, "correct" = identified as AI.

accounting for shared variance. The null effect of human-likeness should be interpreted cautiously: it may indicate that human-likeness influences compliance indirectly through its association with persuasiveness rather than being irrelevant to scam success.

### 4.8. Detection of AI-generated voices: familiarity provides no protection

*Participants struggled to identify AI-generated callers, achieving 70.3% accuracy in voice conditions.* However, participants also frequently misidentified human callers as AI (correctly identifying humans only 24.3–45.8% of the time), suggesting a general heightened suspicion toward callers rather than reliable AI detection ability. Text-based AI detection was barely above chance (53.0% accuracy), while voice recordings allowed significantly better (though still imperfect) detection ($\chi^2(1) = 78.22, p < 0.001$, Cramér's V = 0.17); see Table 3.

*Detection accuracy varied dramatically by AI model.* Relative to OpenAI AVM, participants had significantly worse odds of correctly identifying all other AI voices: Llama FD (OR = 0.50, $p = 0.014$), Gemini (OR = 0.39, $p < 0.001$), Play.AI (OR = 0.38, $p < 0.001$), ElevenLabs (OR = 0.21, $p < 0.001$), and Sesame (OR = 0.14, $p < 0.001$). Sesame was the most difficult to detect, with participants correctly identifying it as AI only 66.3% of the time. Relative to Llama FD (88.6% accuracy), participants had significantly worse odds of correctly identifying Sesame (OR = 0.31, $p < 0.001$) and the ElevenLabs cloned voice (OR = 0.42, $p = 0.002$), but better odds for OpenAI AVM (OR = 1.99, $p = 0.014$).

*Neither general AI familiarity nor voice assistant usage improved detection.* Participants who reported never using AI achieved 54.4% detection of AI voices, compared to 51.2% for those who use AI often/very often, a difference that was not statistically significant ($\chi^2(2) = 0.22, p = 0.896$); see Fig. 6. Similarly, voice assistant usage showed no association with detection accuracy: never-users achieved 53.9% accuracy versus 46.1% for frequent users ($\chi^2(2) = 0.008, p = 0.993$). These null findings suggest that current consumer exposure to AI systems does not confer meaningful protection against AI-powered voice phishing.

*Participants who correctly identified AI voices cited specific conversational artifacts.* For voice conditions, correct detections were associated with noticing "repetitive responses" (OR = 1.38, $p = 0.048$) and "long-winded responses" (OR = 3.80, $p < 0.001$). For text, successful detection correlated with recognizing "unnatural phrasing" (OR = 3.39, $p < 0.001$), "irrelevant responses" (OR = 2.49, $p = 0.019$), and "abrupt sentence transitions" (OR = 2.89, $p = 0.004$). However, these cues were not universally recognized and most participants failed to identify AI even when such artifacts were present (Table 4).

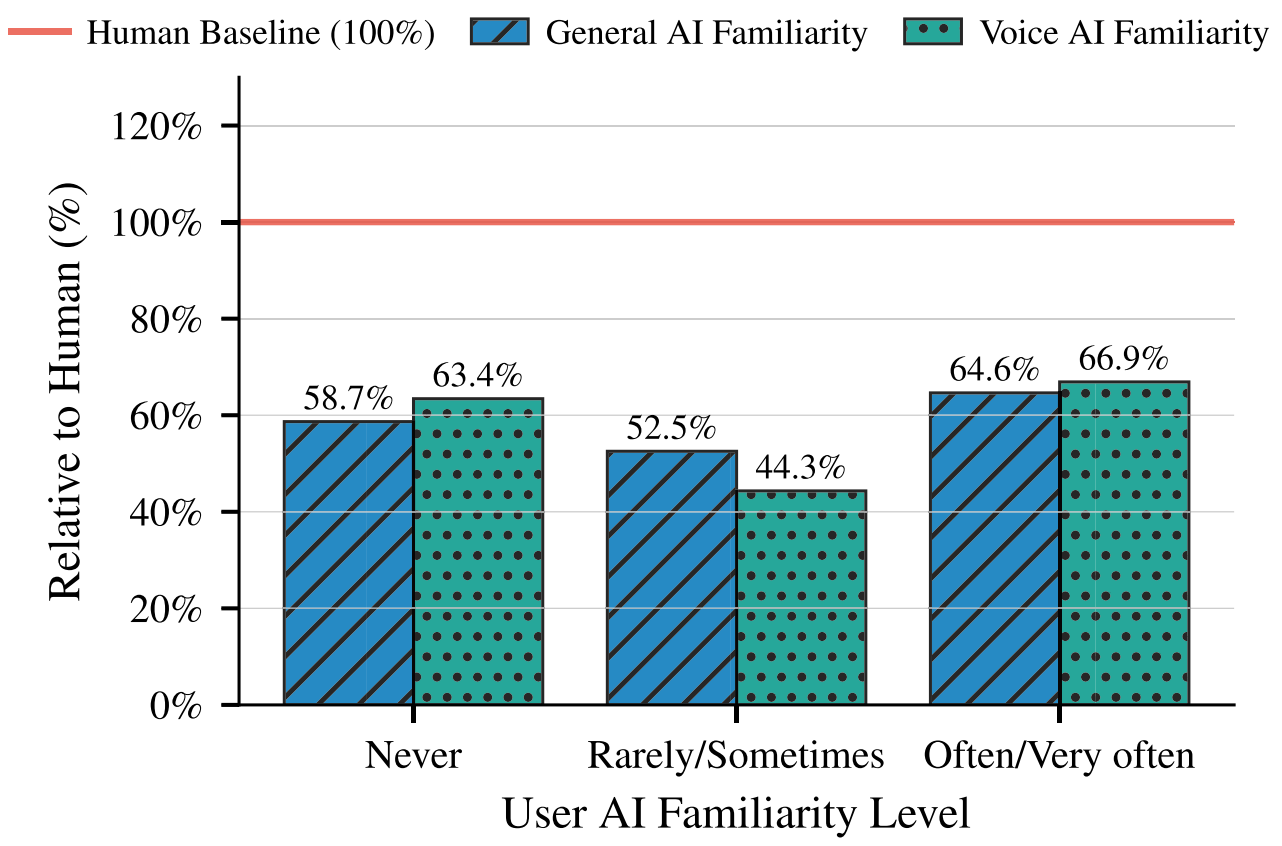

**Fig. 6.** AI misclassification rate relative to human correct classification, grouped by user AI familiarity level. 100% indicates AI voices are misclassified as human at the same rate as human voices are correctly classified.

**Table 4**
Comparison of AI voice models across the three personal distress scams. Sent. = Sentiment, Pers. = Persuasiveness, Trust. = Trustworthiness, H-Like = Human-likeness, Compl. = Compliance.

| Model | Sent. | Pers. | Trust. | H-Like | Compl. (%) |
|---|---|---|---|---|---|
| *Donation scam* | | | | | |
| Llama FD | 3.40 | 2.50 | 2.74 | 2.41 | 20 |
| OpenAI AVM | 3.47 | 2.41 | 2.45 | 2.26 | 26 |
| Gemini | 3.16 | 1.99 | 2.06 | 2.01 | 16 |
| Sesame | 3.14 | 2.21 | 2.22 | 2.79 | 21 |
| Play.AI | 3.61 | 2.56 | 2.61 | 2.65 | 30 |
| ElevenLabs | – | – | – | – | – |
| Human (control) | 3.40 | 2.50 | 2.55 | 3.52 | 23 |
| Transcript (control) | 3.39 | 2.36 | 2.41 | 3.16 | 21 |
| *Police-grandma scam* | | | | | |
| Sesame | 2.87 | 2.35 | 2.19 | 2.78 | 24 |
| Human (control) | 3.05 | 2.46 | 2.30 | 3.57 | 25 |
| Transcript (control) | 3.10 | 2.29 | 2.15 | 3.21 | 20 |
| *Relative-in-distress scam* | | | | | |
| Sesame | 2.97 | 2.44 | 2.33 | 2.81 | 27 |
| ElevenLabs | 3.22 | 2.63 | 2.55 | 3.05 | 32 |
| Human (control) | 3.12 | 2.55 | 2.41 | 3.61 | 29 |
| Transcript (control) | 3.08 | 2.37 | 2.25 | 3.19 | 24 |

## 5. The economics of AI-enhanced vishing

The results presented in Section 4 demonstrate that AI voice systems can approach human-level persuasiveness in scam contexts. We now examine the economic implications of AI-automated vishing, focusing on how automation transforms the cost-benefit calculus for attackers. Traditional vishing has been fundamentally constrained by human labor costs. Each call requires a trained operator who can conduct only one conversation at a time, creating a natural bottleneck that has historically limited the scale of voice-based fraud.

Let $J$ be the set of vishing technologies available (including different AI models and the human option), and consider an attacker using technology $j \in J$ to target an individual $i$ in market $I$. The expected revenue from using $j$ to phish $i$ is:

$$r_j(t, X_i) = m(X_i)p_j(t, X_i)q$$

where $X_i$ is a vector of individual characteristics (such as income, gullibility, or vulnerability profile), $m(X_i)$ is the amount of money that the attacker using $j$ would receive from successfully phishing $i$, $p_j(t, X_i)$ is the probability that $j$ successfully persuades $i$ of their cause, and $q$ is the probability that this persuasion converts into revenue for the phisher. The expected cost for $j$ attempting to phish $i$ is $c_j t$, where $t$ is the time

**Table 5**
Comparison of AI voice models by economic profitability.

| Model | $p_j$ [95% CI] | $c_j$ | Profit [95% CI] |
|---|---|---|---|
| Humans | 0.230 [.205, 0.255] | 2.88 | −27.10 [−27.91, −26.30] |
| Llama FD | 0.120 [.085, 0.155] | 1.30 | −11.71 [−12.84, −10.58] |
| OpenAI AVM | 0.171 [.130, 0.211] | 1.50 | −12.47 [−13.78, −11.16] |
| Gemini | 0.120 [.085, 0.154] | 0.13 | 2.38 [1.25, 3.51] |
| Sesame | 0.152 [.122, 0.182] | 0.33 | 1.03 [0.05, 2.00] |
| Play.AI | 0.171 [.131, 0.212] | 0.90 | −5.25 [−6.57, −3.94] |
| ElevenLabs | 0.369 [.279, 0.459] | 0.75 | 2.97 [0.05, 5.88] |

*Note:* $p_j$ = persuasion probability; $c_j$ = inference cost (or wage); Profit = expected hourly profit ($). CIs omitted for calibrated quantities.

spent vishing and $c_j$ the cost of model inference (or wage rate for humans). Expected profit per vishing attempt is $r_j(t, X_i) - c_j t$.

We obtain $p_j$ from our experiment (averaging over all vishing scams tested by the model) and $c_j$ from posted model prices (and in the case of the human wage rate, calibrate it based on the average US non-farm hourly wage rate of $34.55/hour).[2] This leaves open the question of $q$, the conversion rate from a persuaded individual to an actual payment. To calibrate this quantity, we follow Heiding et al. (2026) and draw on "conversion rates" from the marketing literature as a direct measure of $q$ in legitimate industries. Given that vishers are likely less credible than authentic businesses, we calibrate $q = 0.6\%$ (the lowest observed across real industries).[3] In our baseline analysis, we assume that $m(X_i) = 450$ for all individuals based on Hiya (2024).

Table 5 reveals that using humans to conduct vishing is highly unprofitable at US wages, with an expected loss of $27/hour.[4] On the other hand, the AI models exhibit much greater heterogeneity. Phishers are expected to make negative hourly profits using Llama, OpenAI, and Play.AI, largely driven by the high costs of model inference. On the other hand, Gemini, Sesame, and ElevenLabs are all associated with positive expected profits, in the range of $1-$3 per hour. In the case of Gemini and Sesame, this is due to the very low costs of inference, while in the case of ElevenLabs, the more expensive costs of inference are more than offset by the increase in model quality. This suggests that AI-powered vishing may already be economically profitable for attackers. As models improve in their capabilities, and if costs continue to fall in the industry, we may expect AI-powered vishing to grow more and more profitable.

What happens if vishers can target individuals based on their characteristics (e.g., age, gender, race)? We re-estimate the model under this assumption in Table E.9. This reveals that targeting is not a profitable endeavor across most models (and humans), as the costs of targeting an individual with the right demographic characteristics currently exceeds the returns to tailored persuasion. While personalization may be a powerful tool that may allow AI to influence individuals at increasing levels of granularity, these gains are not currently large enough to justify their costs for attackers.

The development of an automated voice phishing pipeline is costly. At what scale do vishers need to operate in order to justify the cost of developing such a pipeline? Based on our own work in this project, we estimate that the development time for an AI vishing system is roughly 260 h, which corresponds to 5 h per week for 52 weeks. Given that the average hourly wage for a machine learning engineer is roughly $62 per hour (ZipRecruiter, 2026), this amounts to a sunk cost of roughly $16,120 to develop such a tool. For Gemini, Sesame, and ElevenLabs respectively, this implies that the model would have to be continuously vishing for 282, 655, and 226 days in order to justify the costs of developing such a tool.

We note several important limitations of this analysis. Our compliance rates are based on self-reported behavioral intentions rather than observed behavior, likely overstating actual success rates. Additionally, we do not model the time required to convert compliance into actual financial extraction, nor do we account for potential defensive adaptations. Nevertheless, even conservative estimates suggest that AI-automated vishing is economically viable at scales accessible to individual bad actors, not just organized crime syndicates. Furthermore, we implicitly assume that AI-automated vishing is equally scalable as human vishing (with constant returns to scale in both cases): this is likely to underestimate the reach of AI-automated vishing, given that many AI instances can run at the same time and operate many phone lines, as well as work during all hours of the day.

Taken together, this economic analysis suggests that the "automation dividend" (i.e., situations where even small per-call success probabilities translate into large aggregate losses) may be quite empirically relevant. Although the profitability of AI-powered vishing is roughly break-even for attackers across the models surveyed, we note that rapid advances in the technology may likely justify cybersecurity and regulatory action today, especially given the lags that exist in developing tools and laws to counter the rise of these new threats.

## 6. Discussion

Our evaluation of AI-powered voice phishing reveals a complex and rapidly evolving threat landscape where technical sophistication, psychological manipulation, and social engineering converge. Overall compliance rates with AI-generated scams averaged 16.5% (yes or unsure); this conceals dramatic variation across contexts. Personalized, emotionally-charged scams, particularly those using cloned voices, elicited compliance rates up to 36.1%, more than five times higher than generic account support scams. These findings underscore that AI-enabled vishing is not a uniform threat; its risk depends strongly on the emotional context, voice quality, and persuasive strategy employed.

### *6.1. Key findings and implications*

Our findings challenge conventional assumptions about what makes AI-driven deception effective. While voice realism and human-likeness are important, persuasiveness, rather than human-likeness, emerged as the strongest predictor of compliance. Caller sentiment and trustworthiness also contributed significantly, whereas perceived human-likeness did not independently predict susceptibility. This suggests that the psychological narrative and delivery style, not merely acoustic fidelity, determine whether an AI voice succeeds in deceiving targets. When scam context was introduced, perceptions of trustworthiness, sentiment, and persuasiveness declined sharply. Yet this heightened vigilance was insufficient to prevent compliance in emotionally manipulative scenarios, such as the "relative-in-distress" and "grandma" scams, where empathy and urgency overrode suspicion. Importantly, models like Sesame and ElevenLabs achieved near-human or human-level performance in these settings, suggesting that the line between authentic and synthetic persuasion is rapidly blurring. Our detection analysis also found that AI familiarity provided no protection: participants who frequently used AI systems were no better at identifying synthetic voices than those with no AI exposure. Most users failed to notice even when detectable conversational artifacts (e.g., repetition, delayed responses) were present. This indicates that public familiarity with generative AI tools has not yet translated into heightened skepticism or resilience.

[2] To convert costs per minute to costs per vishing attempt, note that the typical conversation in our sample lasts roughly 5 min. Thus, a model can conduct roughly 12 attempts in an hour.

[3] See https://www.invespcro.com/cro/statistics/ and the real estate sector for the low end.

[4] This also suggests that in low and middle income countries, where wages are typically much lower than in the US, this activity may be profitable by humans to perform.

### 6.2. Limitations of the study

This study, while unprecedented in scale, has some limitations that should be acknowledged.

**Experimental Design.** Our approach relied on participants' self-reported behavioral intentions rather than direct behavioral observation. Consequently, compliance rates may not perfectly reflect real-world behavior, particularly in stressful or time-pressured situations. Participants may either underestimate or overestimate their likelihood of falling for such scams. Also, while we include a cloned voice scam in the study, the voice is not known to the participants; thus, what we are measuring is whether a cloned voice and relative in distress scenario can be more effective at scamming individuals compared to a synthetic voice, not whether known cloned voices are subjectively more persuasive for a participant.

**Scenario Design.** To balance ecological validity and control, we selected realistic but not perfectly matched scenarios. For example, the donation scam is not an ideal control for the sister-in-distress condition; a cloned-voice donation request would have been methodologically purer but unrealistic in the current scam ecosystem. Likewise, the cloned voice (researcher's voice) used an Irish accent rather than a North American one, which may have introduced small perceptual differences.

**Audio Quality.** Recordings were intentionally non-professional to reflect realistic scam conditions, but this limited our ability to isolate the effects of voice clarity versus vocal affect. Despite these limitations, the study offers one of the most rigorous population-level estimates to date of how AI voice systems may influence human trust and compliance in fraudulent contexts.

### 6.3. Conclusion

AI-powered voice phishing represents a qualitatively new form of *scalable social engineering*. The results of this study demonstrate that LLMs and voice models can now approximate, and in some cases exceed, human effectiveness in eliciting compliance, especially when emotional or relational cues are present. The threat is measurable and rapidly improving. The implications of these patterns for consumer protection, platform governance, and model release policies require careful consideration. From a target's perspective, a scam is a scam regardless of whether it is conducted by a human or AI. The primary risk of AI-powered vishing is its ability to scale cheaply and effortlessly while maintaining high quality. Our findings underscore the urgent need for cross-disciplinary policy action: (i) *Model governance*: AI developers must move beyond surface-level safeguards and design abuse prevention mechanisms that are robust to removal and circumvention. In practice, however, such protections are technically challenging to implement and often deprioritized amid competitive pressures for rapid innovation and deployment. As a result, many existing mitigations, whether embedded in open-source models or enforced through proprietary usage limits, function as temporary obstacles and are routinely bypassed by malicious actors. This underscores the need for stronger deployment-level monitoring, provenance and auditability mechanisms, and more open research on how to meaningfully constrain abuse at minimal cost to developers. (ii) *Consumer education*: awareness campaigns should focus on recognizing manipulative conversational strategies, rather than detecting AI-generated speech. We ought to promote critical thinking and teach users to resist urgency and emotional pressure, and verify caller identity through independent channels. Education efforts should also include guidance on recovery processes for those who have been victimized, reducing stigma and encouraging reporting. (iii) *Regulatory modernization*: agencies must anticipate the new economics of fraud, in which automation enables attacks to scale at near-zero marginal cost. Regulatory frameworks should incentivize AI developers to prioritize security-by-design while clearly defining accountability and responsibility for downstream harms. One concrete mechanism is risk-based Know Your Customer (KYC) and user verification, which would help deter abuse and support accountability.

In short, AI systems dramatically lower the cost of deception, as shown in Section 5. Defenses must evolve to protect human trust in digital systems and in their users, both human and agentic. AI systems are no longer just tools for productivity and automation. They are potential instruments of persuasion on an industrial scale. Understanding and mitigating their misuse is now an essential frontier in both cybersecurity and public policy. With compliance rates exceeding 30% in some conditions, the potential for harm is substantial, even accounting for the likelihood that self-reported intentions overestimate actual behavior. At scale, even a 5% success rate across millions of automated calls represents a transformative shift in the economics of fraud. The question is no longer whether AI-powered vishing poses a serious threat, but whether policymakers, platforms, and the public will act before the damage becomes irreversible.

## CRediT authorship contribution statement

**Fred Heiding:** Conceptualization, Methodology, Investigation, Supervision, Visualization, Writing – original draft, Writing – review & editing, Project administration; **Claudio Mayrink Verdun:** Formal analysis, Writing – original draft, Writing – review & editing; **Simon Lermen:** Visualization, Writing – review & editing, Writing – original draft; **Andrew Kao:** Investigation, Data curation, Methodology, Writing – original draft, Software, Writing – review & editing; **Vitor Albiero:** Conceptualization, Methodology, Resources, Writing – original draft, Writing – review & editing; **Lauren Deason:** Conceptualization, Methodology, Resources, Writing – original draft, Writing – review & editing; **Irina-Elena Veliche:** Conceptualization, Methodology, Resources, Writing – original draft, Writing – review & editing; **Christine Lehane:** Conceptualization, Supervision, Resources, Funding acquisition, Project administration, Writing – original draft, Writing – review & editing.

## Ethical considerations

This research investigates AI-powered voice phishing, a domain with significant dual-use implications. We conducted an ethics analysis following the Menlo Report principles. Our identified stakeholders, mitigation actions, and justification for conducting and publishing this research are stated below. We identified six stakeholder groups potentially affected by this research: (1) research participants in our survey and interviews who were exposed to scam content, (2) AI model providers including Meta, OpenAI, Google, Sesame, Play.AI, and ElevenLabs, whose products we evaluated in adversarial contexts, (3) organizations impersonated in our scam scenarios, specifically MasterCard and Google, (4) society at large, including potential future victims of AI-powered vishing, (5) the research team, and (6) malicious actors who could potentially misuse our findings.

To protect research participants, all individuals provided informed consent prior to participation, were informed of their rights including the ability to withdraw at any point without penalty or explanation and to request that their data be removed from the project. The survey was conducted by a professional research organization with established protocols for human subjects protection (which will be disclosed provided that the paper is accepted). Because the study involved the secondary analysis of fully de-identified data collected by a professional research organization with established human-subjects protections, and did not involve intervention or interaction by the research team, formal IRB review was not required under U.S. federal regulations governing human subjects research (45 CFR 46.104).

As discussed in Section 3.3, it would be considered unethical to have participants interact directly with an AI in a scam scenario without forewarning them. Our recording-based design allowed us to evaluate susceptibility without exposing participants to these risks and without collecting any actual sensitive information such as credit card numbers or passwords. For this reason, past research using an interactive context has had to (1) make participants aware in advance that they will be interacting with an AI, (2) ask participants not to disclose authentic private information, that is, to fabricate any information shared, and (3) engage in role play, not an authentic scenario. These mitigations are required to protect participants' private information, but they detract from the ability to observe how people would behave and react to the LLM in a real-life circumstance. Our study design, which presented recordings rather than live interactions, allowed us to evaluate susceptibility without exposing participants to these risks.

Our use of mild deception regarding AI involvement, where participants were not informed that callers were AI-generated until after providing evaluations, was methodologically necessary to avoid priming effects, and all participants were debriefed before concluding their session. To mitigate dual-use risks, we deliberately omitted certain operational details. While we report which models achieved highest persuasiveness and which scam types proved most effective, we do not provide complete system prompts in directly executable format. The prompts described in Appendix C are summarized rather than presented verbatim. As discussed in Section 3.2, consumer-facing AI products have appropriate safety mitigations in place, and our recordings were edited to remove these features to represent worst-case scenarios.

The decision to conduct this research was based on our assessment that AI-powered vishing represents an imminent threat, as evidenced by incidents described in Section 2 and the economic analysis in Section 5. Empirical evidence on susceptibility is essential for developing countermeasures and informing policy. The decision to publish was based on our assessment that benefits to defenders and policymakers outweigh incremental risks of informing attackers. The general capability of AI systems to generate scam content is already publicly known. Our contribution provides rigorous susceptibility estimates and identifies psychological mechanisms underlying compliance, findings more useful for defense than offense. Specifically, our finding that persuasiveness rather than human-likeness predicts compliance (Section 4.7) suggests defensive interventions should focus on recognizing manipulative strategies rather than detecting synthetic voices, an insight that primarily benefits defenders. We acknowledge that certain findings, such as high compliance rates for cloned-voice family emergency scams (Section 4.3), could inform more effective attacks. However, withholding these findings would leave defenders less informed while providing minimal protection, given that motivated attackers could conduct similar evaluations independently.

## Declaration of generative AI and AI-assisted technologies in the writing process

During the preparation of this work, the authors used Anthropic's Claude and OpenAI's ChatGPT to assist with data analysis, figure generation, and language refinement. The authors have reviewed and edited all content and take full responsibility for the final text of the publication.

## Funding

This research received no specific grant from any funding agency in the public, commercial, or not-for-profit sectors.

## Declaration of competing interest

The authors declare the following financial interests/personal relationships which may be considered as potential competing interests:

Some authors are affiliated with Meta Platforms, Inc., and Meta researchers contributed to the study design and experimental setup. The authors declare no additional competing financial interests or personal relationships that could have appeared to influence the work reported in this paper. If there are other authors, they declare that they have no known competing financial interests or personal relationships that could have appeared to influence the work reported in this paper.

## Acknowledgments

We thank Rachel Tobac (CEO of SocialProof Security), James Crnkovich (Data Scientist, formerly Meta), Kevin Hannan (Product Manager, formerly Meta), and Aaron Grattafiori (AI Security Researcher, formerly Meta) for their valuable contributions to the design and analysis of this work.

## Data availability

We adhere to open science best practices by documenting our research design, experimental procedures, and analytical methods in sufficient detail to facilitate reproduction, replication, and repetition of our work. As with all human science research, extra care must be taken to ensure methodological validity while strictly complying with ethical and safety constraints. We have taken substantial measures to ensure these criteria are met, including providing comprehensive and transparent descriptions of our experimental procedures and overall setup, including the voice models evaluated, model versions, audio generation pipeline, interaction structure, participant recruitment, scenario design, and outcome measures. We further detail our statistical analyses and reporting choices to enable transparent interpretation of results.

We release aggregated and de-identified quantitative results sufficient to reproduce all reported findings, a complete description of the experimental methodology, including participant demographics, recruitment, study procedures, and scenario design, the full survey instrument and interview discussion guide, and all statistical analysis and data preprocessing code used to generate the reported results and figures. In addition, we provide high-level descriptions of the voice generation approach (see Appendix C), including model classes, versions, deployment constraints, and interaction structure, without releasing operational prompts or implementation details that could be repurposed for real-world abuse.

To support reproducibility, we have created a public repository containing the primary artifacts necessary to evaluate our contributions, including aggregated and de-identified quantitative results, statistical analysis and data preprocessing code, figure-generation scripts, the full survey instrument, interview discussion guide, and qualitative codebooks and thematic summaries. These materials are available at: https://shorturl.at/gC8bH.

Our goal is to establish a rigorous empirical benchmark for evaluating the role of AI in voice-based social engineering and inform mitigation strategies for model developers, platforms, and policymakers, not providing a turnkey vishing capability for anyone to use. To that end, we actively engage with researchers, platforms, and policymakers in this domain and are open to providing additional methodological clarification or controlled access to materials where appropriate to support further academic work.

## Appendix A. Roadmap

This appendix provides the full experimental materials and supplementary analyses referenced in the main text. Complete experimental materials, including the full survey instrument, generation prompts, and interview guide, are available in our anonymized online repository at https://shorturl.at/gC8bH. Appendix B summarizes the survey instrument, including attention checks and outcome measures. Appendix C presents the generation prompts used to create the scam conversations. Appendix D details the qualitative interview methodology, sample composition, and protocol design. Appendix E provides supplementary tables and statistical results supporting the main findings, including baseline model comparisons, descriptive statistics by condition, AI-detection cues, and sample demographics. The economic analysis of AI-powered vishing, examining the cost-benefit calculus for attackers, is presented in Section 5.

## Appendix B. Survey instrument

The survey consisted of three main sections. First, participants received instructions emphasizing the importance of careful attention, as the audio recording (approximately 5 min) or text transcript could only be played or viewed once. An audio test confirmed participants could hear the stimuli. After exposure, an attention check required participants to summarize the conversation content in 1–2 sentences.

Second, participants provided ratings on four dimensions: (1) *caller sentiment* via a single item measuring initial impression (5-point scale from "Very negative" to "Very positive"); (2) *caller persuasiveness* via a single item assessing how convincing the caller was (5-point scale); (3) *caller trustworthiness* via a 9-item scale measuring expertise, credibility, empathy, understanding, friendliness, compelling explanations, trustworthiness, confidence, and importance/urgency ($\alpha = 0.923$); and (4) *caller human-likeness* via the 6-item Partner Modelling Questionnaire using bipolar adjective pairs (warm–cold, personal–generic, empathetic–apathetic, social–transactional, life-like–tool-like, human-like–machine-like).

Participants then indicated their willingness to comply with the caller's request (Yes/No/Unsure), rated their surprise if the caller were revealed to be AI, identified when they first suspected AI involvement, and selected specific conversational elements that prompted suspicion from a checklist. The checklist for audio conditions included items such as unnatural voice, long-winded responses, delayed responses, unnatural speaking rhythm, lack of emotional response, and repetitive responses. The text condition checklist included parallel items adapted for written communication.

Third, participants reported their frequency of AI application use (e.g., ChatGPT, Claude, Meta AI, Gemini) and voice-based AI conversations, rated the ease of following the conversation, indicated any technical issues, and provided demographic information including household income, employment status, industry sector, and job title.

## Appendix C. Generation prompts

All voice-based and text-based scam scenarios used standardized system prompts instructing AI models to adopt specific personas and follow consistent persuasion strategies. The prompts established backstories positioning the AI as legitimate service representatives (e.g., "Lindsay" from Google's account support department in Mountain View, California, or "Jessica" from MasterCard's consumer support services in Purchase, New York), defined clear goals for information extraction, and specified communication styles emphasizing authority, urgency, and guilt-based persuasion.

For the Gmail credential phishing scenario (voice), the prompt instructed the model to: (1) introduce itself and establish urgency regarding suspicious account activity threatening account shutdown; (2) sequentially request confirmation of non-illicit use, email address verification, password confirmation, and a 2FA code sent to the user's phone; (3) refuse alternative remediation options; and (4) maintain an authoritative tone throughout.

For the MasterCard fraud scenario (text), the prompt specified sequential information requests: first and last name, home address with postal code, card number, expiry date, security code, and details of the last legitimate transaction. The model was instructed to warn of $2,000 in suspicious charges and potential liability, refuse alternative verification methods, and maintain urgency throughout.

The donation, grandma, and sister-in-distress scenarios followed similar structures adapted for their respective contexts, with prompts emphasizing emotional appeals, relational framing, and financial urgency appropriate to each scam type.

## Appendix D. Qualitative interviews

**Qualitative Data Analysis.** To complement our quantitative findings with deeper insights into participant reasoning and perceptions, we conducted 12 semi-structured interviews with U.S. adults recruited through AnswerLab, a professional user research firm. As described in Section 3.4, interviews were audio-recorded, transcribed verbatim, and analyzed using thematic analysis following Braun and Clarke's six-phase framework (Braun & Clarke, 2006). The analysis proceeded as follows: the researcher first watched video recordings and read through transcripts to achieve familiarization with the data. Initial codes were generated systematically while reading through transcripts, capturing participant responses to persuasiveness, trust, and perceived authenticity. Codes were then grouped into broader candidate themes corresponding to the primary insight patterns from the data. Themes were reviewed to assess their fit across the interviews, with particular attention to ensuring that themes were not based on single instances but rather recurred meaningfully across participants. Finally, theme names were refined and an analytic narrative was produced. The analysis identified two overarching thematic categories. The first, Human-AI Indistinction, captured participants' difficulty in confidently identifying AI-generated communication. Subthemes included initial assumptions of humanness (where participants attributed robotic qualities to script-reading rather than AI), the role of voice in providing richer evaluative signals compared to text, and lack of awareness regarding current AI capabilities. The second category, Caller Persuasion, revealed that persuasiveness was more strongly connected to conversational content than to the communication medium. Subthemes included how specific content features (such as confidence, plausible details, and personalization) influenced persuasiveness judgments, and how scam literacy shaped participants' susceptibility. Complete thematic codebooks, representative quotes, and detailed subtheme descriptions are available in our repository at https://shorturl.at/gC8bH.

**Sample Composition.** Participants were selected to ensure demographic diversity and varied AI familiarity. The sample was balanced by gender (6 women, 6 men) and distributed across age groups: 3 aged 18–30, 3 aged 31–45, 3 aged 46–60, and 3 aged 61+. All participants had diverse racial and ethnic backgrounds and were regular messaging app users (Instagram Direct, Messenger, WhatsApp) on both iOS and Android systems. AI familiarity ranged from minimal exposure to frequent use of AI tools.

**Interview Protocol.** Each interview lasted on average 61 min and followed a semi-structured format allowing for flexible exploration of emergent themes while maintaining consistency across sessions. The interview compensation was $100. Participants evaluated two scenarios during their session: one audio recording (Gmail phishing via Play.AI voice) and one text conversation (MasterCard phishing via Llama 4 text model). Both scenarios used identical system prompts to those employed in the survey experiment (see Appendix C for full prompts), ensuring consistency between qualitative and quantitative components.

To avoid priming effects, we deliberately withheld information about AI involvement. AI was not mentioned in recruitment materials, consent forms, or session introductions. Participants were told only that they would evaluate "communication scenarios" and share their impressions. To control for order effects, we counterbalanced presentation: half of participants experienced the audio scenario first, and half experienced text first. Only after completing evaluations of both scenarios were participants informed about AI involvement and asked whether they had suspected AI generation.

**Interview Content.** The interview protocol explored four main areas: (1) initial impressions and emotional responses to each scenario, (2) perceived realism and authenticity of the caller/texter, (3) detection strategies and cues used to assess legitimacy, and (4) reasoning processes underlying decisions to trust or distrust the communication. Participants were encouraged to think aloud during scenario evaluation and explain their reactions in detail. Follow-up probes explored specific aspects of voice quality, conversational patterns, and content credibility.

**Data Collection and Analysis.** All interviews were conducted remotely via video conference by neutral third-party interviewers from AnswerLab to minimize researcher bias and social desirability effects. The combination of quantitative and qualitative methods provided both breadth and depth: the survey experiment measured population-level susceptibility and identified patterns across conditions, while interviews illuminated the psychological mechanisms, perceptual processes, and reasoning strategies underlying participant responses.

### *D.1. Qualitative interview insights*

Consistent with research on vocal cues in credibility assessment (Belin et al., 2017; McAleer et al., 2014), participants in the voice condition frequently referenced paralinguistic elements such as tone, pacing, repetition, and assertiveness when evaluating authenticity. Excessive repetition was a common source of skepticism, serving both as a signal of possible malicious intent and as an indicator that the caller might be AI-generated. One participant noted, “I would have hung up right at the beginning and blocked the number, the constant repetition indicates that she’s reading from a script.” Another described the interaction as feeling “robotic” and “like a software that’s just reinforcing what the programmer gave it.” Others attributed unusual vocal characteristics to the caller’s emotional state rather than AI generation: “The tone really like, she started talking a little faster, she started getting repetitive, it was almost like she started getting nervous, like, oh, I’m losing the person here.”

However, suspicion did not automatically translate into rejection. Consistent with truth-default theory (Armstrong et al., 2023), participants often entertained the possibility that the caller was legitimate before revising their judgment. As one participant reflected, “At first, you know, you have to consider, well, this could be legit.” Participants sometimes rationalized robotic or “badgery” qualities as artifacts of call-center scripting rather than evidence of AI generation. Yet when such cues accumulated, skepticism intensified. As one interviewee noted, “It sounded like a human initially, but when it started getting badgery, I actually started to question, is this a person at all?”

Participants were generally skeptical of urgency cues. As one interviewee stated, “At first, it came across as urgency, and of course the agent wants to help you. But on second thought, it reminded me of things I’ve seen on the internet, that if someone is urgently trying to get you to share information, that can be a sign they’re trying to steal it.” Taken together, these qualitative findings reinforce the quantitative results showing that persuasiveness was the strongest predictor of compliance, rather than human-likeness alone. Participants did not rely on categorical judgments of “human vs. AI” when making decisions. Instead, they engaged in post-hoc sensemaking, weighing urgency, confidence, and plausibility, and often defaulted toward provisional trust unless strong countervailing evidence emerged.

## Appendix E. Supplementary tables

This appendix provides comprehensive statistical details supporting the findings presented in Section 4. The tables below offer granular breakdowns of model performance, pairwise comparisons, detection patterns, and demographic characteristics that inform our population-level estimates of susceptibility to AI-powered voice phishing.

### *E.1. Baseline AI model performance in neutral scenarios*

This section provides detailed statistical comparisons and visualizations supporting the baseline analysis reported in Section 4.1. Table E.1 presents pairwise comparisons between AI models on the human-likeness dimension, revealing that Sesame’s advantage over other models (particularly Llama FD, OpenAI AVM, and Gemini) emerges even in benign contexts, suggesting inherent voice quality differences rather than scenario-specific effects.

**Table E.1**
Pairwise comparisons of AI models on caller human-likeness in neutral scenarios.

| Index | Comparison | Diff. ± SE | Sig. | 95% CI |
|---|---|---|---|---|
| Llama | Sesame | $-2.626 \pm 0.895^*$ | 0.035 | [−5.15, −0.10] |
| Llama | Play.AI | $-1.056 \pm 0.890$ | 1.000 | [−3.56, 1.45] |
| Llama | OpenAI AVM | $-0.049 \pm 0.885$ | 1.000 | [−2.54, 2.44] |
| Llama | Gemini | $0.536 \pm 0.899$ | 1.000 | [−2.00, 3.07] |
| Sesame | Play.AI | $1.570 \pm 0.895$ | 0.798 | [−0.95, 4.09] |
| Sesame | OpenAI AVM | $2.577 \pm 0.890^*$ | 0.039 | [0.07, 5.08] |
| Sesame | Gemini | $3.162 \pm 0.904^*$ | 0.005 | [0.62, 5.71] |
| Play.AI | OpenAI AVM | $1.007 \pm 0.884$ | 1.000 | [−1.49, 3.50] |
| Play.AI | Gemini | $1.592 \pm 0.898$ | 0.770 | [−0.94, 4.12] |
| OpenAI AVM | Gemini | $0.585 \pm 0.893$ | 1.000 | [−1.93, 3.10] |

*Note:* $^*p < 0.05$ (Bonferroni-corrected). Starred rows indicate significant mean differences at the .05 level. Bonferroni adjustments were applied to account for multiple comparisons. All results are weighted based on U.S. online population data.

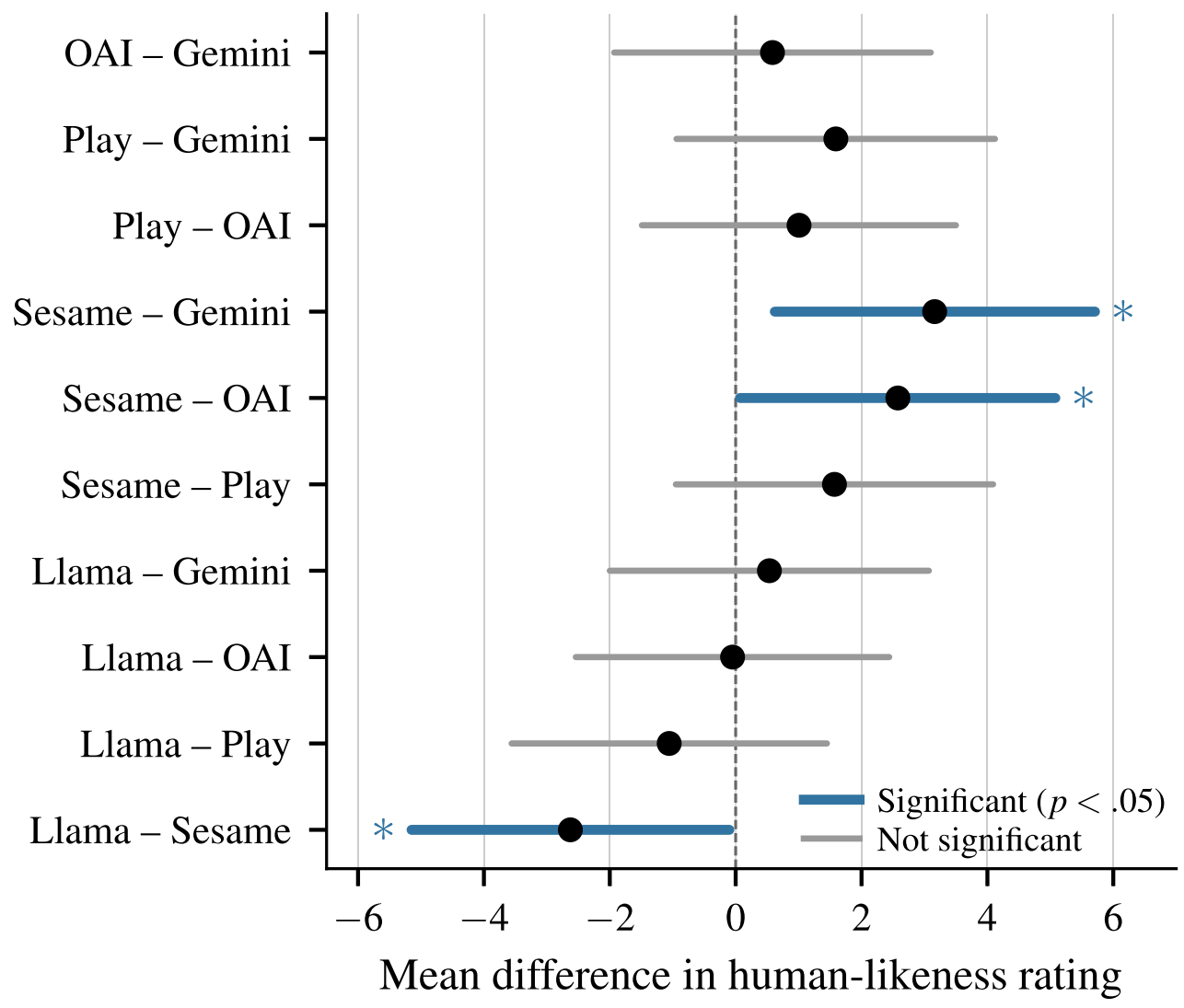

**Fig. E.1.** Pairwise differences in caller human-likeness between AI models in neutral (non-scam) scenarios. Points represent mean differences, and bars show 95% confidence intervals. Blue intervals indicate statistically significant differences ($p < 0.05$).

Fig. E.1 presents mean pairwise differences for each model in caller human-likeness. Sesame was rated as significantly more human-like than Llama FD, OpenAI AVM, and Gemini, but did not differ significantly from Play.AI or the ElevenLabs cloned voice.

Table E.2 extends this baseline analysis by comparing each AI model directly against the human voice control. Fig. E.2 visualizes these comparisons on sentiment and human-likeness dimensions. For sentiment, only Gemini scored significantly lower than the human control ($p = 0.025$). For human-likeness, Sesame was the only model that did not differ significantly from the human control ($p = 0.135$), while all other models were rated as less human-like (all $p < 0.001$).

### *E.2. AI model comparisons in scam scenarios*

Table E.3 presents detailed pairwise comparisons between AI models in scam scenarios involving unknown callers (MasterCard, Gmail, Donation). These results support the analysis in Section 4.4, demonstrating that Sesame significantly outperformed Llama FD across all four

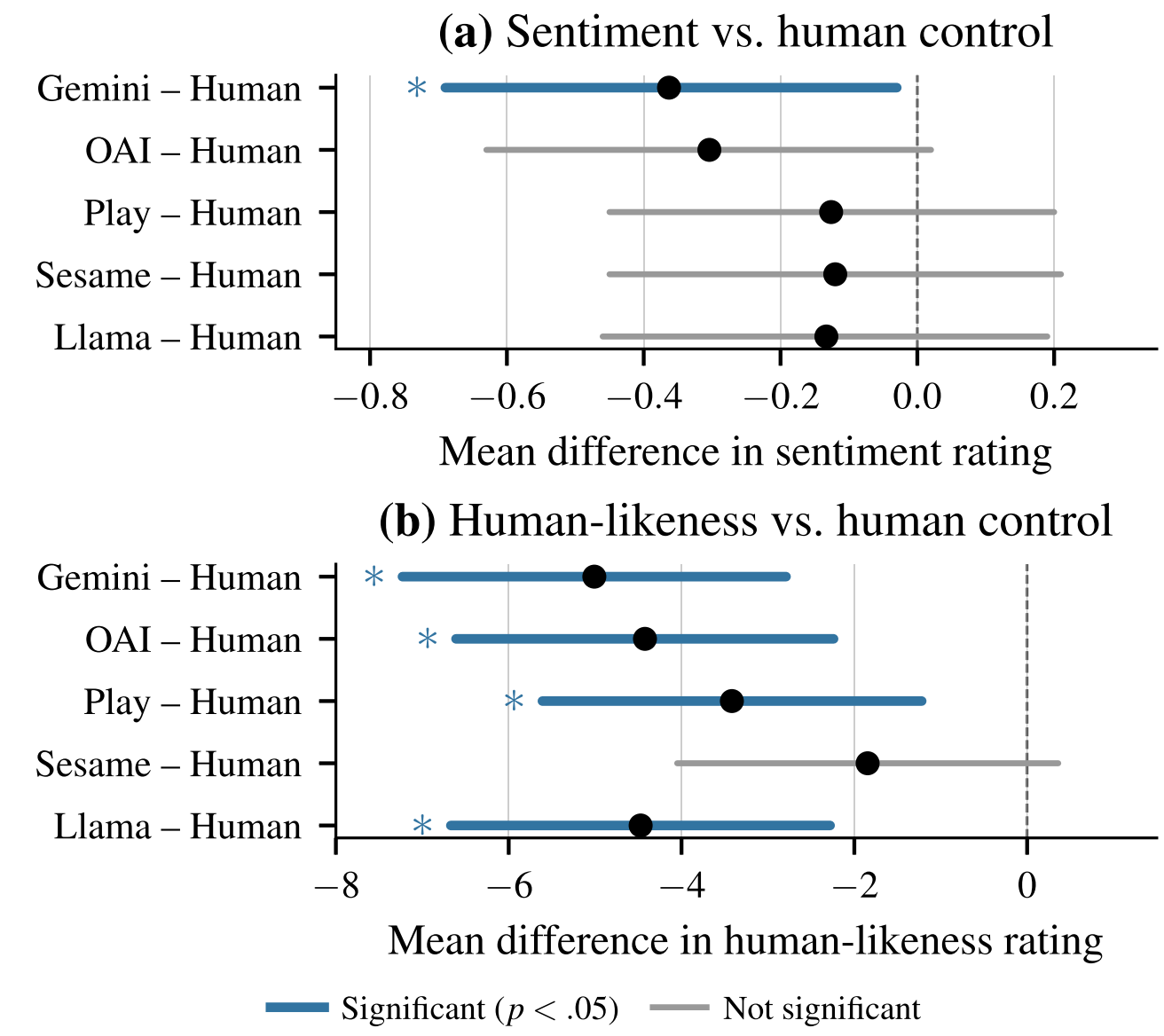

**Fig. E.2.** Comparison of neutral AI models versus a human control voice. Points represent mean differences relative to the human control, with 95% confidence intervals. Blue intervals indicate statistically significant differences ($p < 0.05$).

**Table E.2**
AI models vs human control in neutral scenarios.

| Comparison vs. Human | Mean Diff. | SE | *p* | 95% CI |
|---|---|---|---|---|
| *Caller Sentiment (1–5 scale)* | | | | |
| Llama FD | −0.133 | 0.130 | 0.756 | [−0.46, 0.19] |
| Sesame | −0.120 | 0.131 | 0.827 | [−0.45, 0.21] |
| Play.AI | −0.126 | 0.130 | 0.792 | [−0.45, 0.20] |
| OAI AVM | −0.304 | 0.129 | 0.076 | [−0.63, 0.02] |
| Gemini | −0.363* | 0.131 | 0.025 | [−0.69, −0.03] |
| *Caller Human-Likeness (6–30 scale)* | | | | |
| Llama FD | −4.473* | 0.873 | <.001 | [−6.67, −2.28] |
| Sesame | −1.846 | 0.877 | 0.135 | [−4.05, 0.36] |
| Play.AI | −3.417* | 0.872 | <.001 | [−5.61, −1.22] |
| OAI AVM | −4.423* | 0.867 | <.001 | [−6.61, −2.24] |
| Gemini | −5.009* | 0.881 | <.001 | [−7.23, −2.79] |

*$p < 0.05$ (Dunnett's test, two-tailed). All models compared to human voice control. All results weighted.

perceptual dimensions, while Play.AI showed advantages on trustworthiness and human-likeness.

### E.3. Complete compliance rates by condition

Table E.4 provides comprehensive compliance rates (Yes/No/Unsure) across all AI voice models and control conditions for each scam scenario. These detailed breakdowns support the aggregate findings reported in Sections 4.3 and 4.4, revealing that scenarios combining emotionally resonant content with high-quality voices (e.g., ElevenLabs clone in the sister-in-distress scenario) achieved notably elevated compliance rates.

### E.4. Comprehensive descriptive statistics

Table E.5 provides descriptive statistics (means and s.d.) for all experimental conditions across four key dimensions: sentiment, persuasiveness, trustworthiness, and human-likeness. This table serves as a reference for readers interested in the raw distributional properties underlying the comparative analyses presented throughout Section 5. Researchers conducting meta-analyses or seeking to replicate our findings will find these detailed statistics particularly valuable.

**Table E.3**
AI model comparisons in scam scenarios (Unknown callers).

| Model (vs. Llama FD) | Mean Diff. | SE | *p* | 95% CI |
|---|---|---|---|---|
| *Caller Sentiment (1–5 scale)* | | | | |
| Sesame | 0.322* | 0.101 | 0.005 | [0.08, 0.57] |
| Play.AI | 0.146 | 0.101 | 0.396 | [−0.10, 0.39] |
| OAI AVM | 0.011 | 0.100 | 1.000 | [−0.23, 0.26] |
| Gemini | 0.037 | 0.102 | 0.988 | [−0.21, 0.29] |
| *Caller Persuasiveness (1–5 scale)* | | | | |
| Sesame | 0.261* | 0.096 | 0.023 | [0.03, 0.49] |
| Play.AI | 0.138 | 0.096 | 0.395 | [−0.09, 0.37] |
| OAI AVM | −0.105 | 0.095 | 0.629 | [−0.34, 0.13] |
| Gemini | −0.020 | 0.097 | 0.999 | [−0.26, 0.22] |
| *Caller Trustworthiness (9–45 scale)* | | | | |
| Sesame | 3.574* | 0.674 | <.001 | [1.93, 5.22] |
| Play.AI | 1.934* | 0.674 | 0.015 | [0.29, 3.58] |
| OAI AVM | 0.742 | 0.670 | 0.629 | [−0.89, 2.38] |
| Gemini | 0.309 | 0.684 | 0.975 | [−1.36, 1.98] |
| *Caller Human-Likeness (6–30 scale)* | | | | |
| Sesame | 3.634* | 0.475 | <.001 | [2.48, 4.79] |
| Play.AI | 2.110* | 0.475 | <.001 | [0.95, 3.27] |
| OAI AVM | −0.124 | 0.472 | 0.997 | [−1.28, 1.03] |
| Gemini | 0.207 | 0.482 | 0.979 | [−0.97, 1.38] |

*Note:* *$p < 0.05$ (Dunnett's test, Bonferroni-corrected). All results are weighted. Llama FD serves as the reference group.

### E.5. Detection patterns and cues

Table E.6 breaks down the specific conversational artifacts participants reported when correctly identifying AI-generated callers, complementing the detection analysis in Section 4.8. The table reveals divergent detection strategies across modalities: in voice conditions, "long-winded responses" showed the strongest association with correct detection (OR = 3.80), while in text conditions, "unnatural phrasing" was most predictive (OR = 3.39). These findings suggest that effective detection relies on recognizing mode-specific artifacts rather than universal AI "tells."

### E.6. Aggregated model performance

Table E.7 aggregates performance metrics across all five scam scenarios, providing a bird's-eye view of how different AI models perform on average. This table reveals that ElevenLabs and Play.AI achieved the highest persuasiveness ratings among AI models (2.32 and 2.26 respectively), though still below human controls (2.40). These aggregated comparisons complement the scenario-specific analyses in Sections 4.4 and 4.5, showing that relative model rankings remain fairly stable across contexts.

Table E.8 focuses specifically on the two generic account support scams (Gmail and MasterCard), allowing for direct comparison across models in similar threat scenarios. This table supports the finding in Section 4.4 that Sesame consistently outperformed other models on human-likeness even in generic institutional scam contexts. Notably, compliance rates remained relatively uniform across models in these scenarios (9–15%), suggesting that voice quality differences had limited impact on behavioral outcomes when scam content lacked emotional resonance.

### E.7. Additional analysis on the economics of AI-enhanced voice phishing

The same scam may be more or less appealing to different individuals. We next consider the possibility of using AI to target specific individuals based on their characteristics. To do so, we first estimate the heterogeneous treatment effects on persuasion by model based on a linear model including an individual's age, gender, race, education, and region. For each model, we select the linear combination of

**Table E.4**
Compliance rates: AI voice models and control.

| Model / Type | Scenario | N | No (%) | Unsure (%) | Yes (%) |
|---|---|---|---|---|---|
| *AI Voice Models – Unknown Caller Scams* | | | | | |
| Llama FD | MasterCard | 106 | 93.3 | 3.5 | 3.1 |
| | Gmail | 102 | 87.1 | 4.4 | 8.4 |
| | Donation | 111 | 80.8 | 9.7 | 9.5 |
| Sesame | MasterCard | 117 | 84.9 | 3.7 | 11.4 |
| | Gmail | 116 | 89.2 | 6.2 | 4.5 |
| | Donation | 110 | 72.8 | 12.2 | 15.1 |
| Play.AI | MasterCard | 116 | 91.9 | 6.9 | 1.1 |
| | Gmail | 109 | 87.1 | 10.7 | 2.3 |
| | Donation | 117 | 69.9 | 15.0 | 15.1 |
| OAI AVM | MasterCard | 118 | 88.5 | 9.2 | 2.3 |
| | Gmail | 114 | 84.3 | 9.3 | 6.4 |
| | Donation | 119 | 74.5 | 22.3 | 3.2 |
| Gemini | MasterCard | 114 | 93.7 | 4.1 | 2.2 |
| | Gmail | 106 | 89.1 | 5.9 | 5.0 |
| | Donation | 103 | 82.9 | 12.9 | 4.3 |
| *AI Voice Models – Personal Appeal Scams* | | | | | |
| Sesame (Human script) | Grandma | 94 | 75.2 | 19.5 | 5.3 |
| Sesame (AI script) | Grandma | 113 | 92.8 | 5.4 | 1.8 |
| ElevenLabs Clone | Sister | 107 | 63.9 | 29.6 | 6.5 |
| *Control Conditions – Human Voice* | | | | | |
| Human | MasterCard | 116 | 79.0 | 11.7 | 9.3 |
| Human | Gmail | 111 | 90.7 | 7.1 | 2.2 |
| Human | Donation | 126 | 67.4 | 18.7 | 13.9 |
| Human | Grandma | 110 | 75.9 | 19.8 | 4.3 |
| Human | Sister | 107 | 58.6 | 34.4 | 7.0 |
| *Control Conditions – Text Transcript* | | | | | |
| Transcript | MasterCard | 122 | 87.5 | 8.3 | 4.3 |
| Transcript | Gmail | 110 | 90.7 | 4.3 | 5.3 |
| Transcript | Donation | 118 | 74.4 | 19.7 | 5.9 |
| Transcript | Grandma | 111 | 66.7 | 27.3 | 5.9 |
| Transcript | Sister | 111 | 66.0 | 26.2 | 7.8 |

*Note:* Percentages may not sum to 100% due to rounding. All results weighted.

**Table E.5**
Detailed descriptive statistics by condition.

| Voice | Script | Scenario | N | Sent. *M (SD)* | Pers. *M (SD)* | Trust *M (SD)* | Human *M (SD)* |
|---|---|---|---|---|---|---|---|
| *Human Voice Controls* | | | | | | | |
| Human | Llama 4 | MasterCard | 116 | 3.17 (1.37) | 2.41 (1.37) | 28.81 (9.27) | 17.38 (6.74) |
| | | Gmail | 111 | 2.90 (1.32) | 2.22 (1.17) | 29.18 (8.31) | 17.23 (5.98) |
| | | Donation | 126 | 3.49 (1.20) | 2.83 (1.48) | 32.01 (8.92) | 18.28 (6.75) |
| *Text Transcripts* | | | | | | | |
| Transcript | Llama 4 | MasterCard | 122 | 2.41 (1.20) | 1.89 (1.28) | 22.34 (8.37) | 14.60 (5.33) |
| | | Gmail | 110 | 2.66 (1.30) | 1.90 (1.05) | 22.85 (7.07) | 15.19 (5.49) |
| | | Donation | 118 | 3.54 (1.19) | 2.80 (1.29) | 30.83 (8.91) | 19.84 (6.09) |
| *Llama FD AI Voice* | | | | | | | |
| Llama FD | Llama 4 | MasterCard | 106 | 2.47 (1.17) | 1.89 (1.11) | 21.95 (7.76) | 12.42 (6.09) |
| | | Gmail | 102 | 2.71 (1.28) | 1.95 (1.22) | 24.62 (8.55) | 12.77 (5.66) |
| | | Donation | 111 | 3.36 (1.18) | 2.52 (1.22) | 29.74 (9.33) | 16.37 (6.44) |

*Note:* All results weighted. Sent. = Sentiment, Pers. = Persuasiveness, Trust = 9–45 scale, Human = Human-Like (6–30 scale).

individual characteristics that are predicted to maximize persuasion.[5] We then re-estimate expected hourly profit conditional on successfully targeting this demographic group for vishing.[6] Costs are higher in this category because it is costly to target on demographics: our reference

[5] We also explored an alternate specification where we used the fully interacted set of discretized individual characteristics (a non-parametric approach) rather than linearly adding across marginal treatment effects. We were not powered for such an analysis, even using regularization/shrinkage techniques.

[6] 95% confidence sets are computed using 100 bootstrap iterations.

**Table E.6**
Reasons for suspecting AI-generated caller.

| Detection Cue | Endorsed (%) | OR | 95% CI |
|---|---|---|---|
| *Voice Conditions (AI Scams)* | | | |
| Unnatural voice quality | 52.7 | 1.21 | [0.89, 1.64] |
| Unnatural rhythm | 48.3 | 1.15 | [0.85, 1.56] |
| Unnatural tone | 49.6 | 1.08 | [0.80, 1.47] |
| Repetitive responses | 54.1 | 1.38* | [1.00, 1.89] |
| Long-winded responses | 28.6 | 3.80*** | [2.56, 5.64] |
| Delayed/slow responses | 17.9 | 1.24 | [0.84, 1.84] |
| *Text Transcript Conditions* | | | |
| Unnatural phrasing/lang. | 23.8 | 3.39*** | [2.14, 5.37] |
| Repetitive responses | 38.4 | 1.28 | [0.87, 1.89] |
| Irrelevant responses | 18.7 | 2.49* | [1.16, 5.36] |
| Long-winded responses | 21.3 | 1.64 | [0.98, 2.75] |
| Abrupt sentence starts/ends | 15.8 | 2.89** | [1.42, 5.88] |

*Note:* OR = odds ratio (correct vs. incorrect detection). *$p < 0.05$, **$p < 0.01$, ***$p < 0.001$.

**Table E.7**
Comparison of AI voice models across key metrics (Aggregated data from all five scam scenarios).

| Model | Sent. | Pers. | Trust | Human | Comp. (%) |
|---|---|---|---|---|---|
| Llama FD | 2.83 | 2.09 | 2.09 | 2.10 | 12 |
| OpenAI AVM | 2.87 | 2.09 | 2.11 | 1.99 | 17 |
| Gemini | 2.88 | 2.11 | 2.03 | 2.16 | 12 |
| Sesame | 2.87 | 2.16 | 2.15 | 2.89 | 15 |
| Play.AI | 2.96 | 2.26 | 2.23 | 2.59 | 17 |
| ElevenLabs | 2.95 | 2.32 | 2.29 | 2.84 | 14 |
| Human (ctrl) | 3.04 | 2.40 | 2.34 | 3.45 | 13 |
| Transcript (ctrl) | 3.01 | 2.28 | 2.19 | 3.12 | 12 |

*Note:* ElevenLabs was only tested in the relative-in-distress (clone) scam scenario. All values are means (1–5 scale), except Compliance (%), which reflects "Yes" or "Unsure" responses. Weighted results.

**Table E.8**
Comparison of AI voice models across two account support scams.

| Model | Sent. | Pers. | Trust | Human | Comp. (%) |
|---|---|---|---|---|---|
| *Gmail Scam* | | | | | |
| Llama FD | 2.61 | 1.87 | 1.86 | 1.95 | 9 |
| OpenAI AVM | 2.73 | 1.96 | 2.07 | 2.03 | 14 |
| Gemini | 3.13 | 2.46 | 2.37 | 2.58 | 12 |
| Sesame | 3.18 | 2.43 | 2.59 | 3.32 | 12 |
| Play.AI | 2.74 | 2.23 | 2.29 | 2.61 | 14 |
| ElevenLabs | – | – | – | – | – |
| Human (ctrl) | 2.87 | 2.09 | 2.11 | 3.36 | 13 |
| Transcript (ctrl) | 2.95 | 2.12 | 2.02 | 3.11 | 12 |
| *MasterCard Scam* | | | | | |
| Llama FD | 2.84 | 2.13 | 2.04 | 2.25 | 10 |
| OpenAI AVM | 2.82 | 2.07 | 2.02 | 1.91 | 12 |
| Gemini | 2.79 | 2.08 | 1.98 | 2.13 | 11 |
| Sesame | 2.77 | 2.18 | 2.12 | 2.81 | 14 |
| Play.AI | 2.91 | 2.27 | 2.18 | 2.55 | 15 |
| ElevenLabs | – | – | – | – | – |
| Human (ctrl) | 2.98 | 2.34 | 2.26 | 3.49 | 12 |
| Transcript (ctrl) | 2.93 | 2.22 | 2.14 | 3.15 | 11 |

*Note:* All values are means (1–5 scale) except Compliance (%), which reflects "Yes" or "Unsure" responses. Weighted results.

price of $2.25 per individual is based on the cost YouGov reports to target an individual on the same set of demographic characteristics.[7]

In Table E.9, we report the results. While the persuasion rate $p_j$ is higher across the board,[8] the increase in costs needed for targeting

[7] https://yougov.com/business/products/self-serve-surveys/price-calculator

[8] We note that point estimates can exceed the unit interval when maximizing demographic characteristics are negatively correlated.

**Table E.9**
Comparison of AI voice models by economic profitability (Individually targeted).

| Model | $p_j$ [95% CI] | $c_j$ | Profit [95% CI] |
|---|---|---|---|
| *Targeted* | | | |
| Humans | 0.509 [.050, 0.969] | 24.12 | −45.03 [−59.92, −30.15] |
| Llama FD | 0.478 [.033, 0.923] | 25.70 | −27.12 [−41.54, −12.70] |
| OpenAI AVM | 0.622 [.385, 0.859] | 25.50 | −24.86 [−32.54, −17.18] |
| Gemini | 0.573 [.159, 0.986] | 26.88 | −9.95 [−23.34, 3.45] |
| Sesame | 0.500 [−.296, 0.704] | 26.68 | −14.70 [−21.30, −8.09] |
| Play.AI | 0.937 [.274, 1.60] | 26.10 | −7.43 [−28.92, 14.06] |
| ElevenLabs | 1.56 [1.13, 1.99] | 26.25 | 14.55 [0.54, 28.57] |

*Note:* $p_j$ = persuasion probability; $c_j$ = inference cost (or wage); Profit = expected hourly profit ($). CIs omitted for calibrated quantities.

dominates the analysis, leading to negative expected profits for all models except for ElevenLabs. Thus, although personalization may be an extremely powerful tool that may allow AI to influence individuals at increasing levels of granularity, it does not appear that these gains are currently sufficiently large to justify their costs for attackers.